\documentclass[draft]{agujournal2018}
\usepackage{apacite}
\usepackage{url} 

\usepackage{float}
\draftfalse

\journalname{arXiv}

\begin{document}

%
%


\title{Seasonal, Solar Zenith Angle, and Solar Flux Variations of O$^+$ in the Topside Ionosphere of Mars}

%
%




 \authors{Z. Girazian\affil{1},
 P. Mahaffy\affil{1}, 
 Y. Lee\affil{1,2},
 E. M. B. Thiemann\affil{3}}

\affiliation{1}{NASA Goddard Space Flight Center, Greenbelt, Maryland, USA}
\affiliation{2}{GESTAR, Universities Space Research Association, Columbia, MD, USA}
\affiliation{3}{Laboratory for Atmospheric and Space Physics, University of Colorado Boulder, Boulder, Colorado, USA}



\correspondingauthor{Zach Girazian}{zachary-girazian@uiowa.edu}




\begin{keypoints}
\item The O$^+$ peak forms at a roughly constant atmospheric pressure level on the dayside.

\item The O$^+$ peak density is controlled by the neutral O/CO$_2$ ratio as predicted by photochemical theory.
\item The topside O$^+$/O$_2^+$ ratio decreases with increasing SZA and is highly variable on timescales of days or less.
\end{keypoints}

%
%


\begin{abstract}

	Using observations from MAVEN's Neutral Gas and Ion Mass Spectrometer (NGIMS), we characterize the seasonal, solar zenith angle (SZA), and solar flux dependent variations of the O$^+$ peak and the O$^+$/O$_2^+$ ratio in the topside ionosphere of Mars. We find that the O$^+$ peak is between 220-300 km and forms at a roughly constant neutral atmospheric pressure level of 10$^{(-8.7 \pm 0.4)}$ Pa. The O$^+$ peak altitude also decreases with increasing SZA near the terminator and varies sinusoidally with an amplitude of 26 km over a period of one Mars year in response to the changing solar insolation. The O$^+$ peak altitude reaches a maximum near Northern Winter solstice and  Mars perihelion. The O$^+$ peak density on the dayside has an average value of (1.1$\pm$ 0.5)$\times$ 10$^{3}$ cm$^{-3}$, has no dependence on SZA for SZAs up to $\sim$90$^{\circ}$, and is mainly controlled by the thermospheric O/CO$_2$ ratio as predicted by photochemical theory. Above the O$^+$ peak, the O$^+$/O$_2^+$ ratio in the dayside ionosphere approaches a constant value of 1.1 $\pm$ 0.6, decreases with increasing SZA, and is highly variable on timescales of days or less. We discuss why the O$^+$ peak is different than the main (M2) peak at Mars and why it is similar to the F2 peak at Earth.
   
  \end{abstract}
    ------------------------------------------------------------------------ 
%
%

%


%
%
%
%

\section{Introduction}
\label{sec:intro}
 
	
 \subsection{Motivation and Historical Background}
 
	The O$^+$ and O$_2^+$ layers in the ionosphere of Mars are the primary reservoirs from which plasma escapes the planet \citep{carlsson2006,brain2016}. Characterization of these ion reservoirs is essential for constraining the drivers of their production and the physical processes that control their variability. In this work, we focus on characterizing the spatial and seasonal variations of the O$^+$ layer using observations from MAVEN's (Mars Atmosphere and Volatile EvolutioN) NGIMS instrument (Neutral Gas and Ion Mass Spectrometer) \citep{jakosky2015,mahaffy2015}.
	
	Until recently, the structure and variability of the ionosphere of Mars was studied almost exclusively through analyses of electron density profiles from radio occultation experiments and electron density measurements from the MARSIS radar sounder (Mars Advanced Radar for Subsurface and Ionosphere Sounding)  \citep{gurnett2008,withers2009a,orosei2015}. These analyses have shown that the bulk of the daytime ionospheric plasma is in the M2 layer (also called the F$_{1}$ layer), where the electron density reaches its maximum value at altitudes between $\sim$120-150 km. The basic properties of the M2 layer, such as its peak density and peak altitude, vary with SZA (solar zenith angle) and solar EUV flux in a manner that is consistent with an ionospheric layer in photochemical equilibrium \citep{morgan2008,withers2009a,girazian2013,mendillo2013,fallows2015,vogt2017}. 
		
	To date, only a handful of in situ measurements of the ion composition at the M2 peak have been reported. Two were obtained by the Viking Landers near SZA$\simeq$45$^{\circ}$ \citep{hanson1977} and the others were obtained by MAVEN during one of its ``deep dip'' campaigns near SZA $\sim$90$^{\circ}$ \citep{vogt2017}. These in situ composition measurements showed that the M2 layer is composed of mostly molecular O$_2^+$ ions ($>$85\% of the total ion density). They also showed that the peak of the O$_2^+$ layer coincides with the M2 peak electron density so that the SZA, seasonal, and solar EUV trends of the O$_2^+$ peak are the same as the those of the M2 electron density peak.
	
	At altitudes above the M2 peak, electron densities decrease exponentially with altitude in a region called the topside ionosphere. Unlike the M2 layer, the topside ionosphere is not in photochemical equilibrium because the timescale for plasma diffusion becomes shorter than the timescale for chemical loss near $\sim$180 km \citep{withers2009a,mendillo2011}. The ion composition in the topside ionosphere has also only been measured in situ by MAVEN and the Viking Landers. These measurements have shown that the the O$^+$ density reaches an absolute maximum, or peak, in the topside ionosphere. The Viking Landers measured the O$^+$ peak altitude -- the altitude of the peak O$^+$ density -- at $\sim$230 km and an O$^+$ peak density of $\sim$7 $\times$ 10$^{2}$ cm$^{-3}$. Viking also found that O$_2^+$ and O$^+$ were the dominant species in the topside ionosphere, with O$_2^+$ being more abundant than O$^+$ over most altitudes.
	
        Since the onset of the MAVEN mission, several analyses of O$^+$ densities have been reported. \citet{girazian2017a,girazian2017} presented average altitude profiles of the nightside O$^+$ density and showed that the profiles change during periods of enhanced electron precipitation. \citet{dubinin2018} showed that, above 400 km, O$^+$ densities vary with the solar wind dynamic pressure and interplanetary magnetic field orientation. \citet{thiemann2017a} showed that the O$^+$ peak density increased following the 10 September 2017 solar flare. \citet{benna2015a} showed, using early NGIMS measurements, that O$^+$ densities do not vary much with SZA on the dayside, but have a sharp gradient across the terminator.
    
	In addition, \citet{withers2015a} compared a small subset of NGIMS O$^+$ densities with the Viking Lander O$^+$ densities and found differences between the two datasets. The O$^+$ peak density measured by NGIMS was two times larger than that measured by Viking, and the O$^+$ peak altitude measured by NGIMS was 60 km higher. The authors suggested that these differences might be due to the different season or solar cycle conditions under which the observations were taken. The Viking data came from low solar activity at Solar Longitudes L$_s$ = 97$^{\circ}$ and 118$^{\circ}$, while the NGIMS data came from moderate solar activity at L$_s$=299$^{\circ}$-348$^{\circ}$. These results from \citet{withers2015a} motivate the primary goal of this paper: to characterize the seasonal, SZA, and solar flux variations of the O$^+$ peak density, O$^+$ peak altitude, and topside O$^+$/O$_2^+$ ratio. Such a goal is now attainable because NGIMS has greatly extended its seasonal coverage over the past two years.

 \subsection{Theoretical Background}
      Since observations of the ionospheric composition have historically been limited, many models have been developed with the Viking measurements as their only means of validation \citep{withers2015}. Nevertheless, these models have established the production and loss mechanisms of O$^+$.
      
    As with any ion species, O$^+$ ions must satisfy the continuity equation which, for steady state conditions, is 
	\begin{equation}
	\label{cont2}
	\nabla \cdot([\mathrm{O}^+]\vec{u})= P - L
	\end{equation}
  	
   \noindent{where} [O$^+$] is the O$^+$ density (cm$^{-3}$), $\vec{u}$ is the O$^+$ velocity (cm s$^{-1}$), $P$ is the O$^+$ production rate (cm$^{-3}$ s$^{-1}$), and $L$ is the O$^+$ chemical loss rate (cm$^{-3}$ s$^{-1}$). The two terms on the right hand side of Eq.~\ref{cont2} account for photochemical processes while the term on the left hand side accounts for transport processes.  
   
     In the topside ionosphere, where the O$^+$ peak resides, the production of O$^+$ ions is mainly from photoionization of atomic oxygen \citep{fox2004}. At these high altitudes the thermosphere is optically thin to ionizing EUV photons \citep{fox2004} so the O$^+$ production rate is
    \begin{equation}
	\label{prate}
	P = J_{\mathrm{O}}[\mathrm{O}]
	\end{equation}
   \noindent{where} $J_{\mathrm{O}}$ is the ionization frequency (s$^{-1}$) of atomic oxygen and $[\mathrm{O}]$ is the oxygen abundance (cm$^{-3}$). 
   
   O$^+$ ions are mainly destroyed by the following ion-neutral reaction \citep{fox1996}
     \begin{equation}
     \label{chemloss}
     \mathrm{O}^+ + \mathrm{CO}_{2} \rightarrow \mathrm{O}_{2}^{+} + \mathrm{CO}
     \end{equation}
\noindent{with} a rate coefficient $k_{4} = 1.1 \times 10^{-9}$ cm$^{3}$ s$^{-1}$ \citep{fox2001} so that
   \begin{equation}
	\label{lrate}
	L = k_{4}[\mathrm{O}^+][\mathrm{CO}_2]
	\end{equation}
 \noindent{where} $[\mathrm{CO}_{2}]$ is the carbon dioxide abundance (cm$^{-3}$). 
 
 Combining Eqs.~\ref{cont2}-\ref{lrate}, the continuity equation governing the O$^+$ density at altitudes above $\sim$200 km in the Martian thermosphere is
     \begin{equation}
	\label{main}
	 \nabla \cdot([\mathrm{O}^+]\vec{u})   = J_{\mathrm{O}}[\mathrm{O}] - k_{4}[\mathrm{O}^+]\mathrm{[CO}_2].
	\end{equation}

Equation~\ref{main} states that the O$^+$ density depends on transport processes, such as vertical and horizontal flows, the solar EUV flux, and the abundances of thermospheric O and CO$_2$. If photochemical equilibrium is satisfied such that the transport term is small compared to the chemical loss term ($\nabla \cdot([\mathrm{O}^+]\vec{u}) << L$), Eq.~\ref{main} can be solved for the O$^+$ density:
     \begin{equation}
	\label{pce}
	 [\mathrm{O}^{+}] = \frac{J_{\mathrm{O}}}{k_4}\left[\frac{\mathrm{O}}{\mathrm{CO}_2}\right].
	\end{equation}
\noindent{Under} photochemical equilibrium, then, the O$^+$ density depends only on the solar flux and the O/CO$_2$ ratio in the thermosphere. All of the quantities in Eq.~\ref{pce} are measured by MAVEN instruments. This will allow us to test if this equation provides an accurate prediction of the O$^+$ peak density.%


   

\subsection{Objectives}

	Our objectives are to quantify the seasonal, SZA, and solar flux variations of the O$^+$ peak density, O$^+$ peak altitude, and topside O$^+$/O$_2^+$ ratio using MAVEN NGIMS observations that cover a full Martian year. In Sections~\ref{sec:Data}-\ref{sec:averaging} we discuss the MAVEN data and our methodology. In Section~\ref{sec:vert} we show how the O$^+$ peak density and peak altitude vary with SZA, EUV flux, and season, and test if the O$^+$ peak density satisfies Eq.~\ref{pce}. In Section~\ref{sec:ratio} we quantify the O$^+$/O$_2^+$ ratio in the topside ionosphere. In Sections~\ref{sec:discussions} and \ref{conclusions} we discuss our results and present our conclusions.

\section{MAVEN Data}
\label{sec:Data}

	MAVEN's 4.5 hour elliptical orbit has a periapsis altitude that usually varies between $\sim$145-165 km, although it has been periodically lowered to $\sim$125 km for week long ``deep dip campaigns''. During each periapsis pass MAVEN spends $\sim$20 minutes below 400 km allowing instruments to sample the bulk of the ionosphere along a non-vertical trajectory. The periapsis of MAVEN precesses slowly in latitude and SZA so that, from one orbit to the next, the spacecraft encounters nearly the same latitudes and SZAs (usually within 0.5$^{\circ}$). The geographic longitudes from one orbit to the next, however, differ by $\sim$65$^{\circ}$ due to Mars' planetary rotation. Over the long time span of the mission this slow precession allows sampling of the ionosphere at a wide range of latitudes, SZAs, L$_{s}$, and heliocentric distances.

	Our analysis focuses on observations of thermal O$^+$ ions from the MAVEN NGIMS instrument \citep{mahaffy2015}, which obtains in situ density measurements of neutrals and ions during each periapsis pass \citep{benna2015a,mahaffy2015a}. NGIMS has two primary observing modes -- one to measure only neutrals and one to measure both neutrals and ions. The observing mode is usually toggled after every periapsis pass so that ion measurements are obtained every other orbit (9 hours apart).  
	
\begin{figure}[ht!]
 \centering
\includegraphics[width=1.0\textwidth]{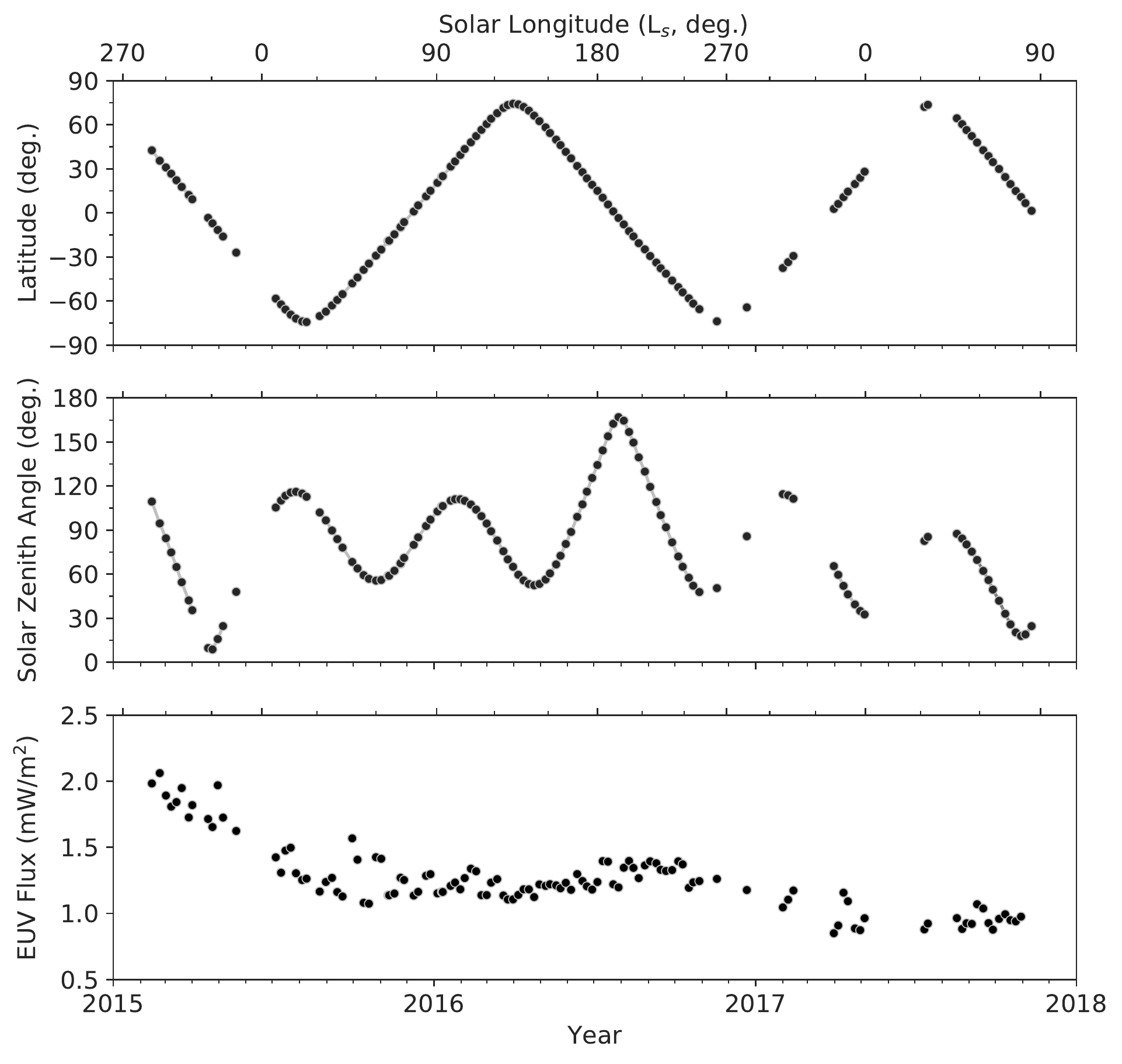}  
\caption{Distribution of the data used in our analysis. The top two panels show, respectively, the geographic latitude and SZA at periapsis as a function of year and L$_s$. The bottom panel shows the EUV flux (as defined in the text) at Mars for the same time period. The data points in each panel correspond to every $\sim$30 MAVEN orbits, although some data gaps are present.
}
\label{dataselection}
\end{figure}

	We use Level 2 NGIMS O$^+$ and O$_2^+$ measurements that have been calibrated using the spacecraft potential (the calibrated densities are in the Level 2 files). We also filter out any unreliable density measurements that were obtained when the spacecraft potential exceeded -4 volts. The filtered dataset contains $\sim$2150 periapsis passes from 11 February 2015 through 14 November 2017. Note that the NGIMS data obtained prior to 11 Feb. 2015 is not used here because it has not yet been calibrated using the spacecraft potential (data version v07\_r03). Figure~\ref{dataselection} shows a representation of how the data is distributed in time, L$_s$, geographic latitude, and SZA. The data span more than a full martian year and cover most latitudes and SZAs.
	
	In addition to ion densities, we also use neutral densities and solar EUV fluxes in our analysis. For the neutral data we use Level 2 closed source NGIMS measurements (CO$_2$, O, N$_2$, CO, and Ar) that were obtained during the same periapsis passes as the filtered ion dataset. We restrict the neutral data to measurements obtained on the inbound legs of each periapsis pass because data from outbound legs are not yet fully calibrated. We also sum the densities weighted by their masses to derive the total mass density (kg km$^{-3}$) at every measurement location.
	
	For the solar flux we use Level 3 data from MAVEN's Solar EUV Monitor (EUVM) \citep{eparvier2015,thiemann2017}. The Level 3 data provide daily-averaged EUV solar spectra derived using the FISM-M empirical spectral model. We integrate the daily-averaged spectra over all wavelengths less than 92 nm (the longest wavelength that can ionize CO$_2$ and O) to get a single-number representation of the solar EUV flux at Mars which we call $F$ \citep{girazian2013,girazian2015b}. The EUV flux, shown in Fig.~\ref{dataselection}, varies by a factor of $\sim$2 throughout the dataset. This variation is due to the changing heliocentric distance of Mars during its yearly orbit around the Sun, as well as the weakening of solar activity during the declining phase of Solar Cycle 24. For historical context, the solar activity level of the dataset covers solar moderate to solar minimum conditions \citep{lee2017}. 
    
   In addition, we also use the FISM-M spectra to calculate the ionization frequency of atomic oxygen, $J_\mathrm{O}$. This is calculated by integrating the spectra over all wavelengths less than 92 nm, and weighting the flux in each wavelength bin, ($F_{\lambda}(\lambda$)), with the atomic oxygen ionization cross section, $\sigma_{i}$, such that $J_\mathrm{O}$ = $\sum F_{\lambda}(\lambda)\sigma_{i}(\lambda)$).

\section{Averaging Procedure}
\label{sec:averaging}

\subsection{Ion Data}

\begin{figure}[ht!]
\centering
\includegraphics[width=1.0\textwidth]{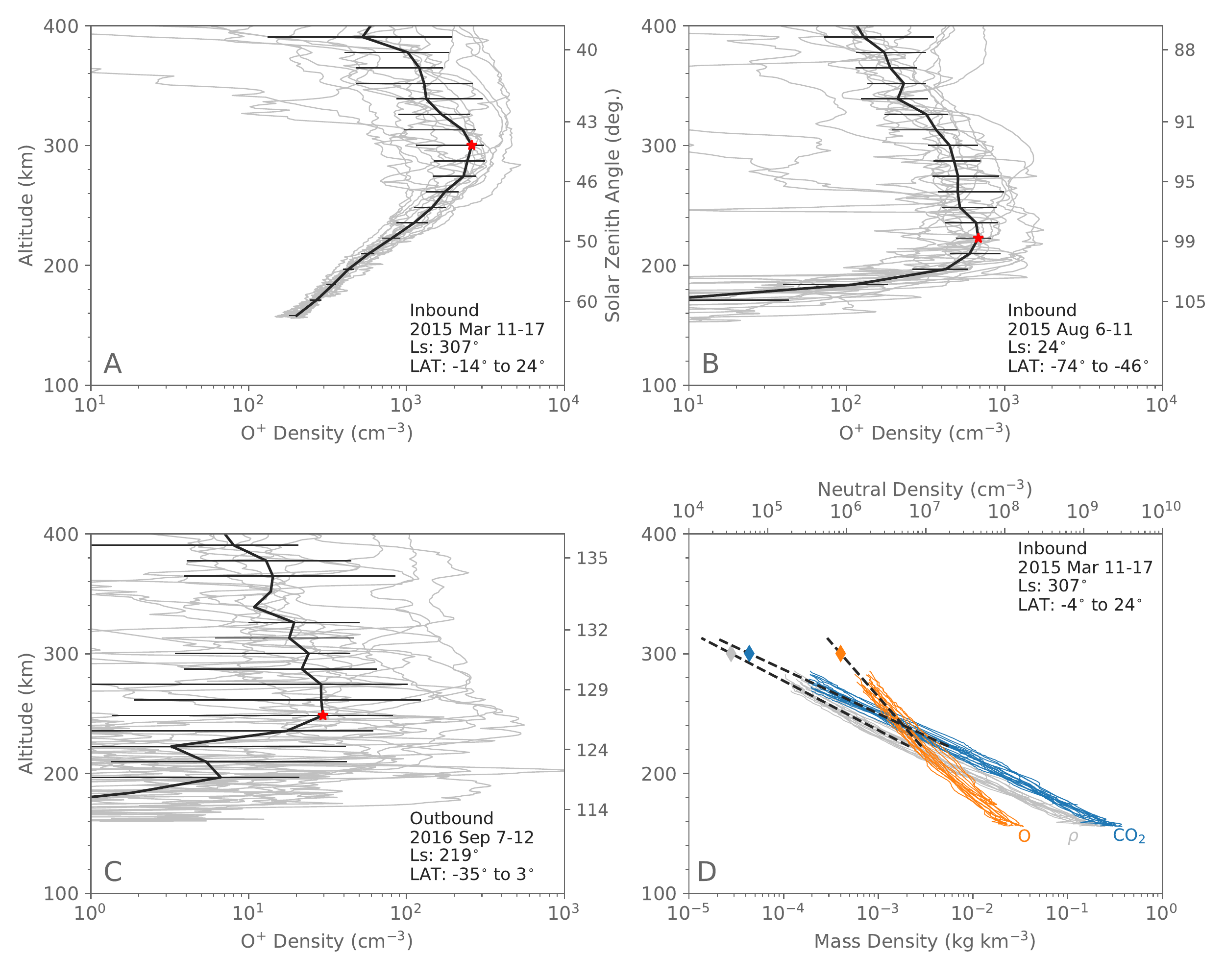}  
\caption{\textbf{A-C)} Each panel demonstrates our procedure for computing an average O$^+$ profile. The gray lines show NGIMS O$^+$ data from $\sim$15 periapsis passes during either inbound or outbound. The black line shows the average profile, computed by separating the data into 13 km-wide altitude bins and finding the median density in each bin. The error bars show the 25\% and 75\% O$^+$ density quartiles in each altitude bin. The right vertical axis shows the SZA range of MAVEN during the periapsis pass. The red star marks the peak altitude and peak density of the average profile. Note that Panel C has a different horizontal axis scale. \textbf{D)} Similar to Panels A-C but for the neutral CO$_2$, O, and mass density ($\rho$). The neutral data are from the same orbits that were used to make the O$^+$ profile shown in Panel A. The dashed lines are an exponential fit to the data (see text) and the diamonds are the density values at the O$^+$ peak.}
\label{fits}
\end{figure}

	Our analysis focuses on the vertical structure of the O$^+$ layer, its peak properties, and their variations with SZA, season (L$_{s}$), and solar flux. Since these variations have timescales of weeks or longer, we average the O$^+$ data to remove variations that occur on shorter timescales. Our averaging procedure, described below and illustrated in Figure~\ref{fits}, produces altitude profiles of the O$^+$ layer by averaging data from $\sim$30 consecutive MAVEN orbits.
	
	To make the average profiles we first take all the O$^+$ data from 30 consecutive periapsis passes. Then, treating the inbound and outbound data separately, we divide the data into 13 km wide altitude bins and find the median O$^+$ density in each bin. We chose a bin size of 13 km because it is large enough to produce relatively smooth median profiles with well-defined peaks, but is small enough to capture major features in the O$^+$ vertical structure. Additionally, we chose to average over 30 periapsis passes after trial and error. The 30 passes provide near-complete coverage in geographic longitude. Furthermore, the 30 passes cover $\sim$5.5 Martian days, which is long enough to allow for averaging over short-term variations in upstream solar wind conditions, but short enough so that MAVEN's periapsis location does not change significantly over 30 passes.
    
    Figure~\ref{fits} shows three examples of this procedure. We iterate through the entire dataset with this averaging procedure to make one inbound and one outbound average profile every 30 periapsis passes. Note that these are not running averages: O$^+$ measurements from each periapsis pass are used in (at most) one and only one average profile. The spacing between adjacent average profiles is roughly $\sim$5.5 Martian days. 
    
    In total, 232 average O$^+$ profiles are produced, 122 of which are from the dayside with SZA$<$80$^{\circ}$. The latitudes and SZAs of the data (at periapsis) used to derive each average profile are shown in Fig.~\ref{dataselection}. Note that each average profile is computed with data that span a wide range of geographic longitudes so that the effects of crustal magnetic fields and atmospheric tides, if present, are averaged out.

	Since NGIMS obtains ion data every other orbit (Sec.~\ref{sec:Data}), each average profile is computed using O$^+$ densities from $\sim$15 periapsis passes (30 consecutive orbits, only half of which have NGIMS ion measurements). In practice though, the number of periapsis passes used to compute each average profile varies between 8 and 19 because of data gaps, changes in the NGIMS observing mode, and our filtering of the data (Sec.~\ref{sec:Data}). 
	
	MAVEN travels along a non-vertical trajectory through the atmosphere so that during a typical orbit the SZA along MAVEN's trajectory changes by $\sim$25$^{\circ}$ between 300 km and 150 km (see Fig.~\ref{fits}). Since our profiles are computed by averaging data in the vertical direction, they may have an altitude bias. Namely, the upper part of the average profile is derived from measurements at one SZA, while the lower part of the average profile is derived from measurements at a different SZA. At small SZAs these horizontal variations have only a small effect on the derived average profile because the O$^+$ density changes gradually with SZA \citep{benna2015a,girazian2017}. 
	  
	
	When near the terminator, however, the O$^+$ density has a steep gradient with SZA  \citep{benna2015a,girazian2017} that will affect the derived average profiles. An example of an average profile derived from data obtained while MAVEN crossed the terminator is shown in Fig.~\ref{fits}B. In this example, the SZA changes from 87$^{\circ}$ to 105$^{\circ}$ over the averaging range. This clearly introduces an altitude bias in the average profile and it is likely that the average profile does not represent the true vertical structure of the ionosphere near the terminator. To help evade this problem, in our analysis we focus primarily on average profiles derived from dayside orbits when the SZA never exceeded 80$^{\circ}$. 
	
	
	Another source of uncertainty in the average profiles is a consequence of MAVEN's orbital precession, which causes the spacecraft to encounter slightly different SZAs, altitudes, and latitudes on adjacent orbits (Sec.~\ref{sec:Data}). The error introduced by this precession is small because, when grouping data from 30 consecutive orbits, the SZA within each 13 km altitude bin changes by only $\sim$10$^{\circ}$. 
	
	From every average profile we extract the O$^+$ peak altitude, peak density, and their uncertainties (also shown in Fig.~\ref{fits}). We define the uncertainty in the peak density as the standard deviation of the O$^+$ densities that are in the same 13 km altitude bin as the peak. The uncertainties in the peak densities as a fraction of the peak density ($\sigma_{N_{\mathrm{max}}}/N_{\mathrm{max}}$) are $\sim$0.35 on the dayside and $\sim$3.0 on the nightside. The example in Fig.~\ref{fits}C shows that the uncertainties in the nightside peak densities are so large because nightside ion densities vary tremendously from orbit-to orbit \citep{girazian2017,girazian2017a}. For the peak altitude, we estimate the uncertainty to be $\pm$13 km in every profile, which is equivalent to two altitude bins -- the minimum needed to resolve a peak.
	
	We also repeat this entire procedure to compute average profiles of the O$_2^+$ density and the O$_{2}^{+}$/O$^+$ ratio. These average profiles are used in Sec.~\ref{sec:ratio} of our analysis. We do not extract peak properties from the average O$_2^+$ profiles because the O$_2^+$ peak is usually below MAVEN's periapsis altitude \citep{vogt2017}.

\subsection{Neutral Data}

	We also average neutral NGIMS neutral density measurements in a similar manner. This allows us to assign an average neutral density profile to each average O$^+$ profile. We use our O$^+$ averaging procedure on the neutral CO$_2$, O and total mass density data, only using data from orbits that also have O$^+$ measurements. Then, we fit each average density profile with a single scale height exponential which allows us to extrapolate the neutral density profiles to higher altitudes as needed. Fig.~\ref{fits}D shows an example, namely the average CO$_2$, O and mass density profiles derived using data from the orbits that were used to make the average O$^+$ profile shown in Fig.~\ref{fits}A. 

    The altitude range which is used to fit the exponential changes for each average profile. It is set by finding the maximum altitude in which there are many NGIMS observations and then fitting the data that is between the upper altitude and 60 km below the upper altitude. For typical dayside conditions the scale heights fits are usually at $\sim$200-250 km.

	Our exponential fits to the data are necessary because the NGIMS neutral density measurements mostly do not extend up to the altitude of the O$^+$ peak (see Fig.~\ref{fits}D). By extrapolating the density measurements to higher altitudes, we obtain a measurement of the CO$_2$, O, and mass density at the O$^+$ peak. To estimate the uncertainty of each extrapolated value, we use the spread in the observational data at 210 km after scaling it to the O$^+$ peak altitude. In particular, the uncertainty in each extrapolated value is set to $\sigma_{extrap} = \left(\sigma_{210} / \tilde{x}_{210} \right) x_{extrap}$ where $\sigma_{210}$  is the standard deviation of the data in the bin centered at 210 km, $\tilde{x}_{210}$ is median value of the data in the bin centered at 210 km, and $x_{extrap}$ is the extrapolated value.
    
\subsection{EUV Data}
	We also assign an average EUV flux and ionization frequency to each O$^+$ profile. To derive these we use the median EUV flux (or ionization frequency) from the groups of orbits used to make each average O$^+$ profile.

\section{Analysis}
\label{sec:analysis}

\subsection{Vertical Structure of the O$^+$ Layer}
\label{sec:vert}

\begin{figure}[ht!]
\centering
\includegraphics[width=1.0\textwidth]{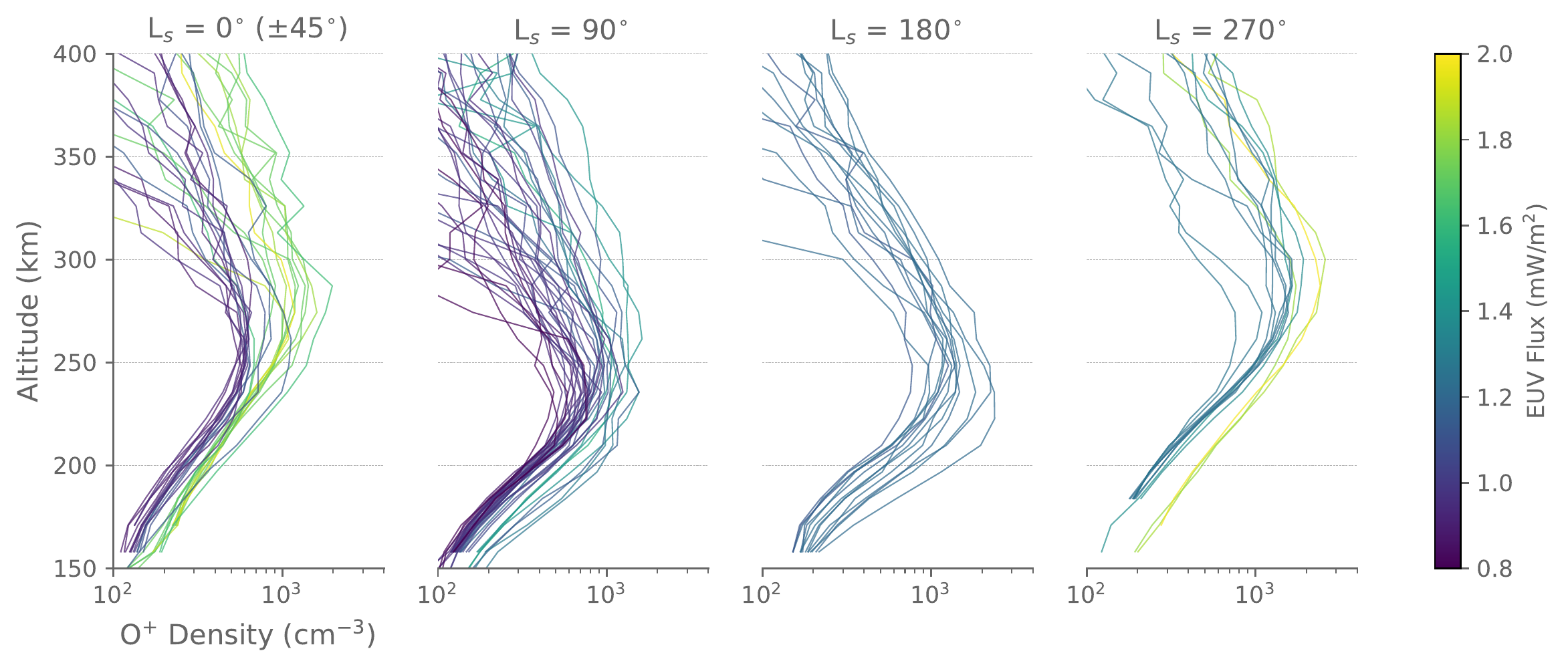}  
\caption{The average O$^+$ profiles from the dayside ionosphere (SZA$<$80$^{\circ}$) grouped by season and colored according to EUV flux. The four panels show profiles from $\pm$45$^{\circ}$ centered on L$_s$ = 0$^{\circ}$, 90$^{\circ}$, 180$^{\circ}$, and 270$^{\circ}$.}
\label{averages}
\end{figure}

	Figure~\ref{averages} shows that the vertical structure of the dayside O$^+$ layer is the same at all seasons and EUV levels encountered thus far during the MAVEN mission, with densities that peak between $\sim$220-300 km and decrease exponentially above and below the peak. This vertical structure is consistent with the O$^+$ measurements made by the two Viking Landers \citep{hanson1977} and is generally consistent with predictions from numerical models that include transport processes such as vertical diffusion \citep{withers2015}. 	
	
	Comparing profiles from within each panel of Fig.~\ref{averages} shows that the O$^+$ density varies more at higher altitudes than at lower altitudes. This variability might be expected given that at higher altitudes the neutral atmosphere is more tenuous and so transport processes play an important role in controlling the O$^+$ density structure \citep{fox2009a,mendillo2011,chaufray2014}. The transport processes may, in part, be controlled by interactions between the ionosphere and the solar wind \citep{ma2004,dubinin2018}. By contrast, at lower altitudes the O$^+$ density is in photochemical equilibrium and its variations -- driven primarily by the neutral atmosphere and EUV flux (Eq.~\ref{pce}) -- are much smaller.
	
	Although the shape of O$^+$ layer is the same across all seasons and EUV levels, there are some features in the profiles that differ when comparing the profiles across the four panels in Fig.~\ref{averages}. First, the typical O$^+$ peak altitude varies from panel to panel, which suggests a seasonal variation. Second, the O$^+$ densities at all altitudes increase with increasing EUV flux. In the next two sections we quantify these trends focusing on the O$^+$ peak.

\subsection{SZA and EUV Flux Variations of the O$^+$ Peak}
\label{peakvars}

\begin{figure}[H]
 \centering
\includegraphics[width=.60\textwidth]{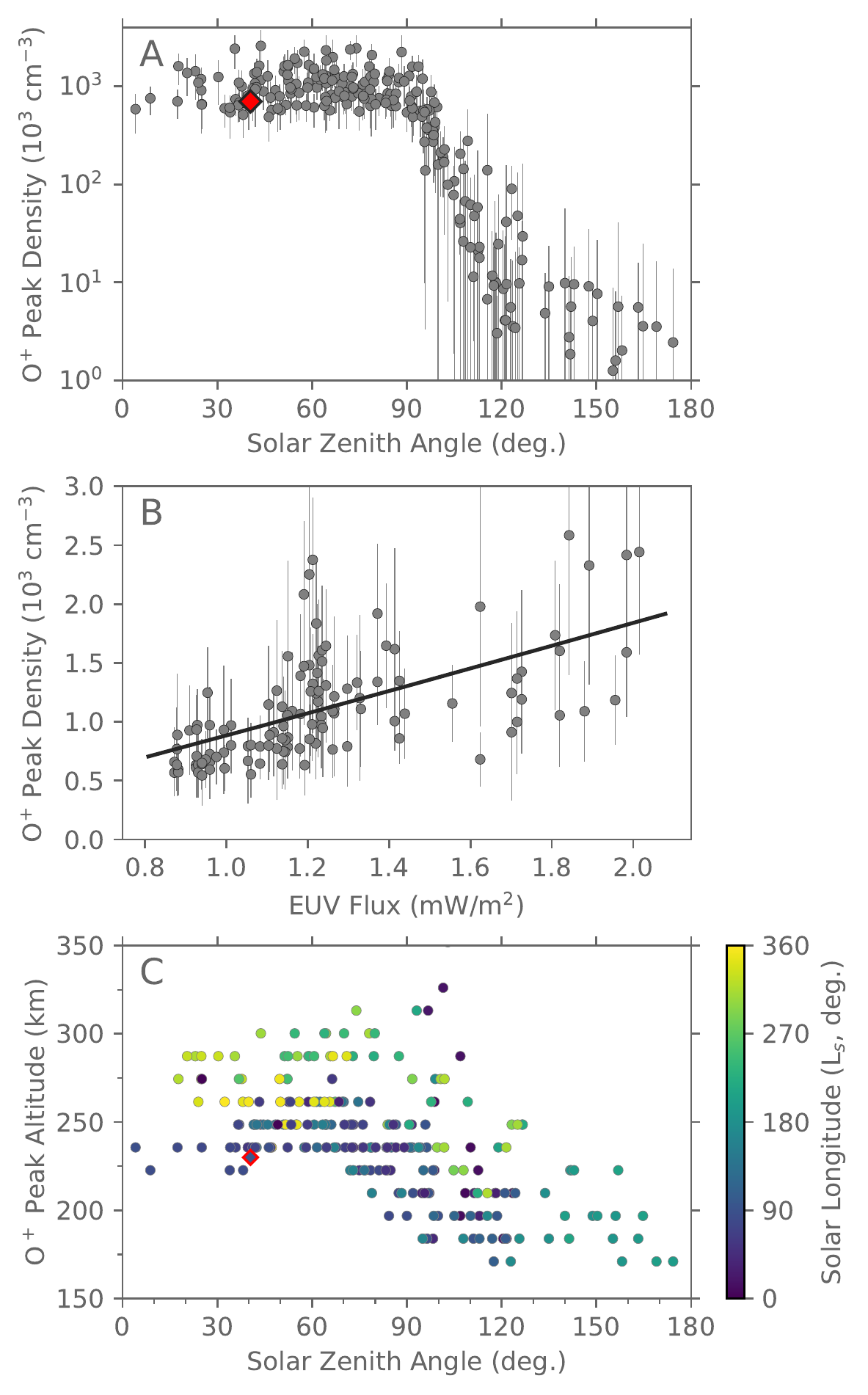}  
\caption{\textbf{A)} O$^+$ peak densities as a function of SZA. \textbf{B)} Dayside O$^+$ peak densities (SZA$<$80$^{\circ}$) as a function of EUV flux. The black line is a fit to Eq.~\ref{nmfit}. \textbf{C)} O$^+$ peak altitudes as a function of SZA and L$_s$. The red diamonds in Panels A and C show the peak density and peak altitude measured by the Viking 1 Lander. }
\label{nmax}
\end{figure}

	In this section we show how the O$^+$ peak density varies with SZA and EUV flux, and how the O$^+$ peak altitude varies with SZA. Figure~\ref{nmax}A shows the O$^+$ peak densities as a function SZA. The dayside peak densities are approximately constant with SZA up to SZA$\simeq$90$^{\circ}$, but there is significant variability. For SZA$<$90$^{\circ}$ the peak density varies between 0.5 -- 2.6 $\times$ 10$^{3}$ cm$^{-3}$ with an average value of 1.1 $\pm$ 0.5 $\times$ 10$^{3}$ cm$^{-3}$. Near the terminator, at SZAs between 90$^{\circ}$-120$^{\circ}$, the peak densities decrease with increasing SZA as the EUV flux decreases while approaching the optical shadow. On the nightside past 120$^{\circ}$ SZA, the peak densities are small ($<$10 cm$^{-2}$) and variable due to the short chemical lifetime of O$^+$ in the presence of CO$_2$, highly variable electron precipitation rates, and perhaps day-night plasma transport \citep{girazian2017,girazian2017a}.

    The dependence of the O$^+$ peak density on EUV flux is shown in Fig.~\ref{nmax}B for SZA$<$80$^{\circ}$. The peak densities can be fit to a power-law function of the form 
    \begin{equation}
    \label{nmfit}
    N_{\mathrm{max}} = N_0{F}^{k}
    \end{equation}
    where $F$ is the EUV flux, $N_{\mathrm{max}}$ is the peak density, and $N_0$ is the peak density when $F$=1.0 mW m$^{-2}$. The best-fit parameters are $N_0$ = 0.89 $\pm$ 0.12 $\times$ 10$^{3}$ cm$^{-3}$ and $k$ = 1.1 $\pm$ 0.4. The exponent $k$ is nearly 1.0 implying that the peak density varies close to linearly with the EUV flux. There is, however, significant scatter around the best fit line and a large uncertainty in the derived exponent. This implies that photoionization is not the only process that controls the O$^+$ peak density, consistent with the prediction of Eq.~\ref{main}, which states that the peak density should also depend on plasma transport and neutral densities. We will explore the role of neutral densities in Sec.~\ref{peakdensity}.

    Furthermore, the observational coverage of the NGIMS dataset is incomplete in latitude, local time, and season, making it difficult to robustly determine how the O$^+$ peak density depends solely on the EUV flux. 
 
    Figure~\ref{nmax}C shows the O$^+$ peak altitudes as a function of SZA. The dayside peak altitude varies between $\sim$220-300 km and is roughly constant with SZA up to SZA$\simeq$70$^{\circ}$. The peak altitude then decreases with increasing SZA across the terminator reaching $\sim$160-200 km on the nightside. This is in contrast the main peak of the ionosphere where the peak altitude increases with increasing SZA \citep{withers2009a}. Consequently, the vertical separation between the O$^+$ and O$_2^+$ layers is larger at the subsolar point than near the terminator.
    
    At any given SZA the dayside peak altitudes span a wide range of values. The wide range of peak altitudes is clearly caused by a trend with $L_s$. This seasonal variation will be examined more closely in the next Section.
    
	For historical context, the O$^+$ peak density and peak altitude observed by the Viking 1 Lander is plotted with a red diamond in Figs.~\ref{zmax}A and C. Both quantities are consistent with the MAVEN observations.

\subsection{Seasonal Variations}
\label{seasons}

\subsubsection{O$^+$ Peak Altitude}
	
\begin{figure}[ht!]
 \centering
\includegraphics[width=0.75\textwidth]{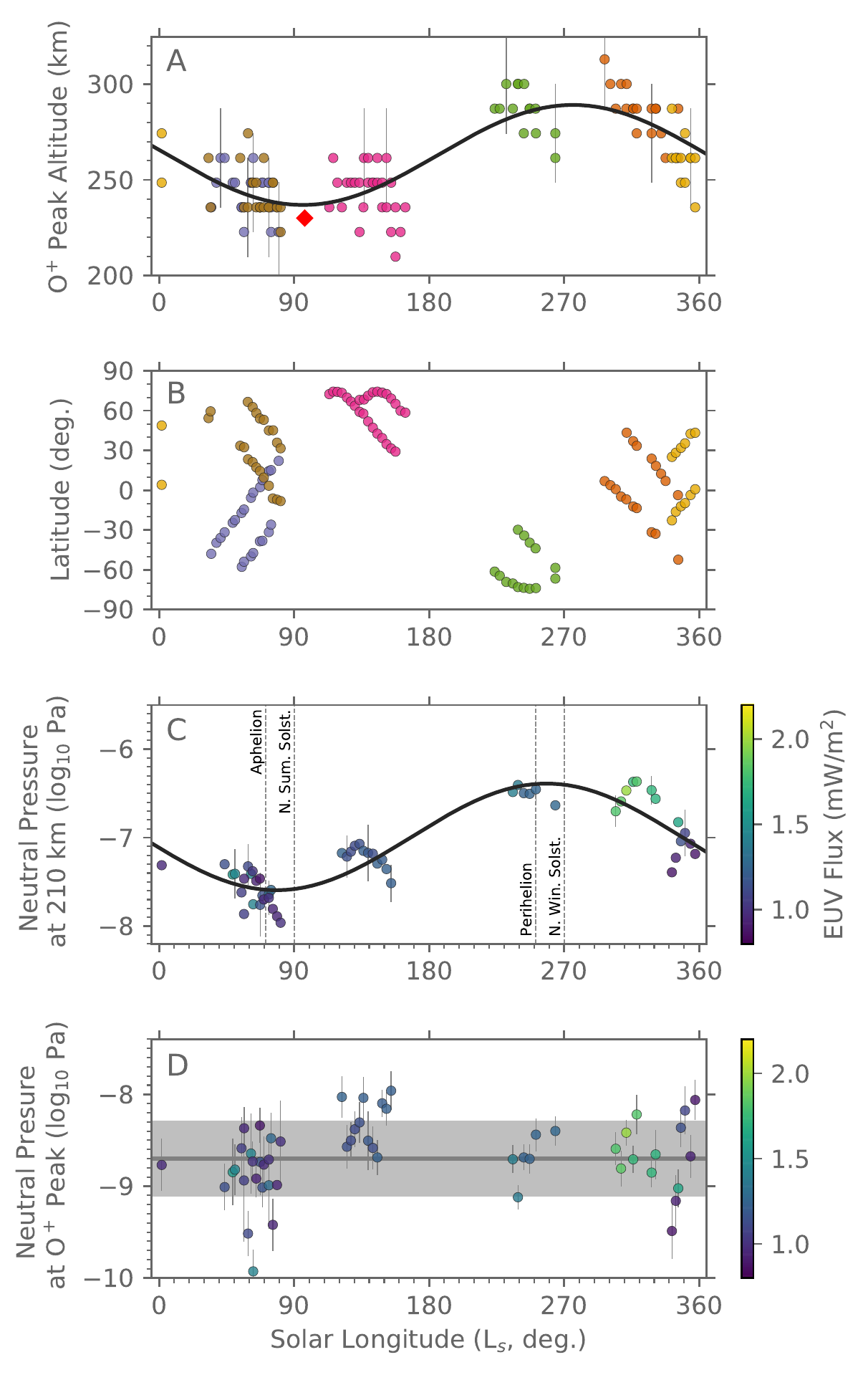}  
\caption{\textbf{(A)} The seasonal variation of the O$^+$ peak altitude. The red diamond shows the Viking 1 measurement. The colors group data from similar time periods: Mar.-May 2015 (orange), Sep.-Dec. 2015 (purple), Mar.-May 2016 (pink), Oct.-Nov. 2016 (green), Mar.-May 2017 (yellow), and Sep.-Nov. 2017 (brown). \textbf{(B)} The geographic latitudes of the peak altitudes. \textbf{(C)} The seasonal variation of the neutral atmospheric pressure at 210 km. \textbf{(D)} The neutral atmospheric pressure at the O$^+$ peak. The thick grey line marks the average pressure and the grey shaded region marks $\pm$1$\sigma$. The black curves in (A) and (C) show the fit to Eq.~\ref{zmfit}. The colors in (C) and (D) show the EUV flux during each observational period. All panels show only dayside data with SZA$<$80$^{\circ}$.  }
\label{zmax}
\end{figure}

    Fig~\ref{zmax}A shows the dayside (SZA$<$80$^{\circ}$) peak altitudes as a function of L$_s$. The colors in the plot group data from similar time periods. Although there is significant scatter in the peak altitudes at any given L$_{s}$, Fig~\ref{zmax}A suggests that the O$^+$ peak altitude has a seasonal variation. When analyzing this seasonal trend we must be cautious because the data have incomplete L$_s$ coverage and are obtained at different geographic latitudes during each season (Fig.~\ref{zmax}B). Keeping in mind these shortcomings, we can quantify the seasonal variation of the peak altitude using a function of the form 
	\begin{equation}
	\label{zmfit}
	    h_{\mathrm{max}} = A\cos{\left[ \frac{\pi}{180^{\circ}} (\mathrm{L}_s + B) \right] + C}
	\end{equation}
\noindent{w}here $h_{\mathrm{max}}$ is the peak altitude, $A$ is the amplitude of the seasonal variation, $B$ is the phase, and $C$ is the average value of $h_{\mathrm{max}}$. The best-fit parameters derived using least squares minimization are $A$ = 26 ($\pm$ 5) km, $B$ = 84$^{\circ}$ ($\pm$ 12$^{\circ}$), and $C$ = 263 ($\pm$ 4) km. The reduced chi-squared of the fit is 1.1 which suggests that Eq.~\ref{zmfit} is a good model for fitting the data.

	The best-fit value A = 26 km implies that the peak altitude varies by 52 km throughout one Mars year. The best-fit value B = 84$^{\circ}$ implies that the peak altitude reaches a minimum at L$_{s}^{min}$ = 96$^{\circ}$ and a maximum at L$_{s}^{max}$ = 276$^{\circ}$. These maximum and minimum values nearly coincide with the Northern Summer (L$_s$ = 90$^{\circ}$) and Northern Winter (L$_s$ = 270$^{\circ}$) solstices. We caution, however, that a precise determination of L$_{s}^{min}$ and L$_{s}^{max}$ is not possible given the incomplete coverage of the MAVEN data (Fig.~\ref{dataselection}). In particular, there are several L$_s$ gaps in the data, including near L$_{s}^{min}$ and L$_{s}^{max}$.  
	
	Furthermore, Eq.~\ref{zmfit} is not a perfect model for fitting the data: the peak altitudes are scattered around the fit and there are trends within some of the individual time-period groupings not captured by the fit. These shortcomings are a result of Eq.~\ref{zmfit} being a simple model that only captures L$_{s}$ variations. The peak altitude may also be dependent on other factors such as geographic latitude. As shown in Fig.~\ref{zmax}B, the observations at each L$_s$ are from different latitudes and the latitude slowly evolves in each time-period grouping. In addition, the data are nonuniformly distributed with respect to solar EUV flux. Nevertheless, the peak altitude appears to have a sinusoidal seasonal variation with a minimum near L$_s$ = 90$^{\circ}$ and a maximum near L$_s$ = 270$^{\circ}$. The sinusoidal variation is supported by the observations near L$_s$ = 70$^{\circ}$ and 350$^{\circ}$. During both of these time periods there are repeated observations separated by one Mars year, and in both cases the peak altitudes are well fit by Eq.~\ref{zmfit}.

    

    
	The seasonal variation of the O$^+$ peak altitude is driven by seasonal variations in the thermospheric pressure. This is demonstrated with Figure~\ref{zmax}C, which shows that the thermospheric pressure at 210 km, $P_{210}$, varies with L$_s$ in a similar way as the O$^+$ peak altitude. The thermospheric pressure was calculated using $\rho gH$ where $\rho$ is the mass density, $g$ is the gravitational acceleration, and $H$ is the neutral scale height of $\rho$. The correlation coefficient between the O$^+$ peak altitudes and $P_{210}$ is 0.82. Fitting $\log_{10}\left( \frac{P_{210 km}}{1.0 \mathrm{Pa}}\right)$ to Eq.~\ref{zmfit} yields best-fit parameters $A$ = 0.60 ($\pm$ 0.08), $B$ = 102$^{\circ}$ ($\pm$ 8$^{\circ}$), and $C$ = -7.0 ($\pm$ 0.06). The reduced chi-squared of the fit is 1.1 which suggests that Eq.~\ref{zmfit} is a good model for fitting the data.
	
	The best-fit value B = 102$^{\circ}$ implies that $P_{210}$ reaches a minimum at L$_{s}^{min}$ = 78$^{\circ}$ and a maximum at L$_{s}^{max}$ = 258$^{\circ}$. These maximum and minimum values are slightly out of phase with the corresponding values for the O$^+$ peak altitude (although they are comparable within the uncertainties), but they nearly coincide with aphelion (L$_s$ = 71$^{\circ}$) and perihelion (L$_s$ = 251$^{\circ}$) of Mars.

    
       
	This seasonal variation in the thermospheric pressure is strongly coupled to the heliocentric distance of Mars, which varies between 1.38 AU and 1.67 AU during a Mars year. Due to the inverse-square law, this results in the solar insolation at Mars varying by $\sim$45\% throughout one Mars year which affects the thermospheric pressure in two ways. 
    
    First, variations in the solar flux at infrared wavelengths cause temperatures in the lower atmosphere to vary, leading to inflation and contraction of constant pressure levels in the thermosphere \citep{bougher2000a,galindo2013}. Second, variations in the solar flux at EUV wavelengths cause temperatures in the upper atmosphere to vary, thereby changing the thermospheric scale height and pressure at a given altitude \citep{jain2015,bougher2017,thiemann2018}. Thermospheric pressure levels can also be affected by the annual CO$_2$ polar cap cycle \citep{wood1992}.

    The O$^+$ peak altitude responds by moving up and down in tandem with these pressure variations, a phenomenon that is predicted by models \citep{bougher2000a,fox2004,galindo2013,chaufray2014}. Fig.~\ref{zmax}D explicitly shows that the thermospheric pressure at the O$^+$ peak (SZA$<$80$^{\circ}$) is essentially constant (although we note that the pressure at the O$^+$ peak is slightly higher near L$_s$$\sim$140$^{\circ}$). The average pressure $\log_{10}[\rho g H/1.0 \mathrm{\ Pa}])$ at the O$^+$ peak is $-8.7 \pm 0.4$. This is consistent with Fig. 3 of \citet{thiemann2017a}, which shows NGIMS O$^+$ densities as a function of atmospheric pressure for a handful of MAVEN orbits.

    

    
	Finally, the O$^+$ peak altitude observed by the Viking 1 Lander is plotted with a red diamond in Fig.~\ref{zmax}A. The peak altitude observed by Viking is consistent with the seasonal variations found in the MAVEN data, suggesting that these trends are stable over decade-long timescales.

 \subsubsection{O$^+$ Peak Density}
 \label{peakdensity}
 
\begin{figure}[H]
 \centering
\includegraphics[width=0.7\textwidth]{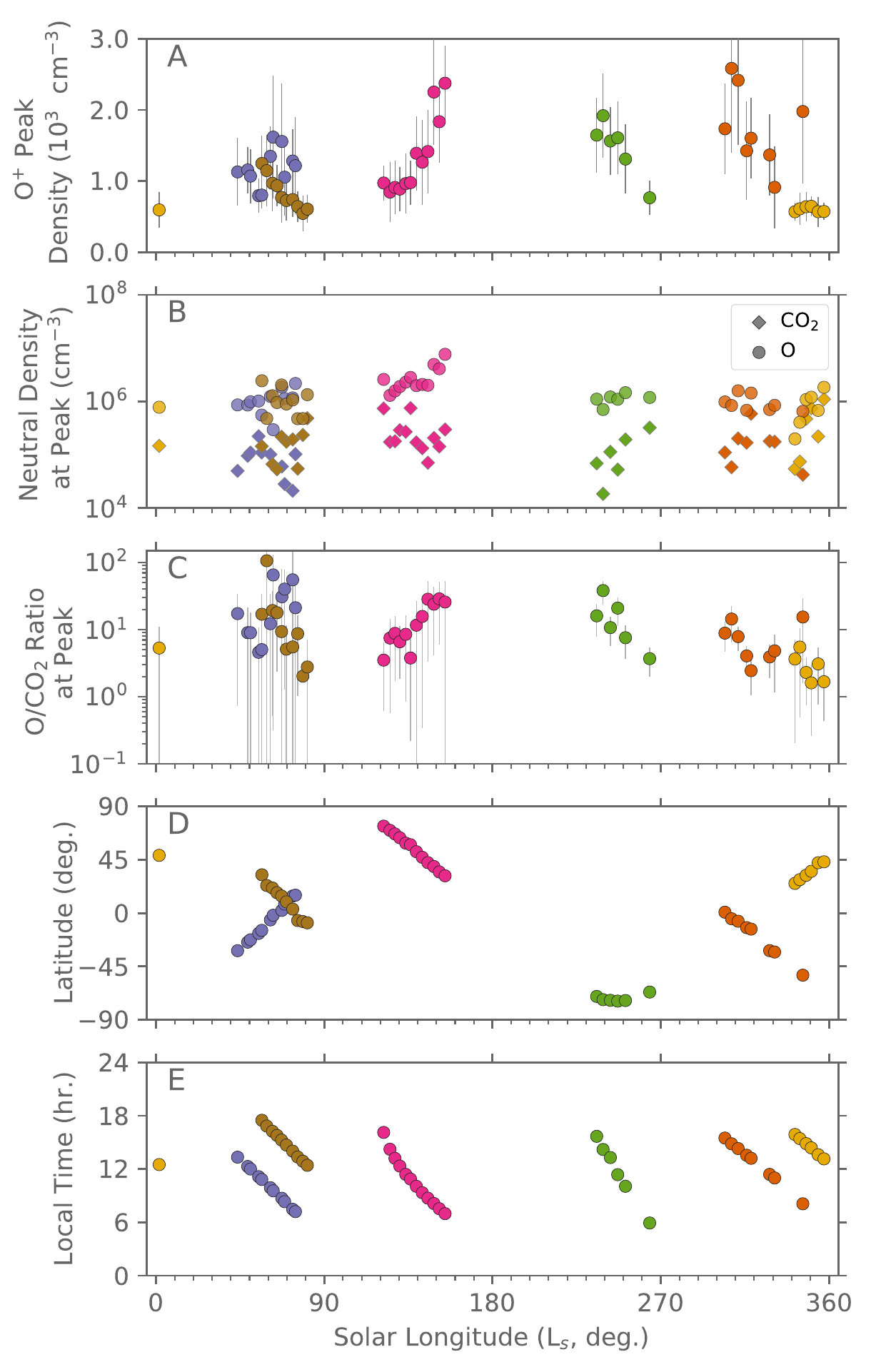}  
\caption{\textbf{A)} The dayside O$^+$ peak density as a function of L$_s$ (SZA$<$80$^{\circ}$). \textbf{B)} The neutral O and CO$_2$ density at the O$^+$ peak as a function of L$_s$. \textbf{C)} The neutral O/CO$_2$ ratio at the O$^+$ peak as a function of L$_s$. \textbf{D)} The latitudes of the observations. \textbf{E)} The local times of the observations. In each panel the colors group data from similar time periods as described in Fig.~\ref{zmax}. }
\label{nmseason}
\end{figure}

    The variability of the O$^+$ peak density with L$_s$ is shown in Fig.~\ref{nmseason}A. The colors in the plot group data from similar time periods. The most striking trends in the plot are    
    within these groupings of data. For example, the peak densities between L$_{s}$=120$^{\circ}$-180$^{\circ}$ (pink circles) 
 steadily increase with L$_s$. This steady increase in the O$^+$ peak density coincides with a steady increase in the neutral O density and the O/CO$_2$ ratio at the peak, which are shown respectively in Figs.~\ref{nmseason}B and C. The O density increased during this period as MAVEN's periapsis precessed from a geographic region where the O abundance at the peak was relatively low (LAT=75$^{\circ}$, LST=16 hr), to a region where it was relatively high (LAT=30$^{\circ}$, LST=7 hr). Meanwhile, the CO$_2$ density was relatively constant resulting in a net increase in the O/CO$_2$ ratio. This dependence of the O$^+$ peak density on the O/CO$_2$ ratio is consistent with photochemical equilibrium predictions (Eq.~\ref{pce}). The increase in the O abundance near dawn is a result of dayside produced O being transported nightward by global circulation patterns. This transport mechanism creates dawn and pre-dawn bulges in the densities of light atmospheric species such as He and O, but not heavy species such as CO$_2$ \citep{bougher2015,elrod2017}.   

Similar trends are seen in two other time periods: L$_{s}=$230$^{\circ}$-270$^{\circ}$ (green circles) and  L$_{s}=$50$^{\circ}$-90$^{\circ}$ (brown circles). During these periods the peak O$^+$ density decreased with increasing L$_s$ as MAVEN's periapsis precessed in latitude and local time such that there was a net decrease in the O/CO$_2$ ratio. 

These O$^+$ peak density trends with L$_s$ are clearly not driven solely by seasonal effects. Instead, they are a consequence of MAVEN encountering the O$^+$ peak at different latitudes and local times as its periapsis slowly evolves. The O$^+$ peak density trends are controlled by the local O/CO$_2$ ratio for a given latitude and local time, as is predicted by the photochemical equilibrium (PCE) equation for the O$^+$ density (Eq.~\ref{pce})  

\begin{figure}[ht!]
 \centering
\includegraphics[width=\textwidth]{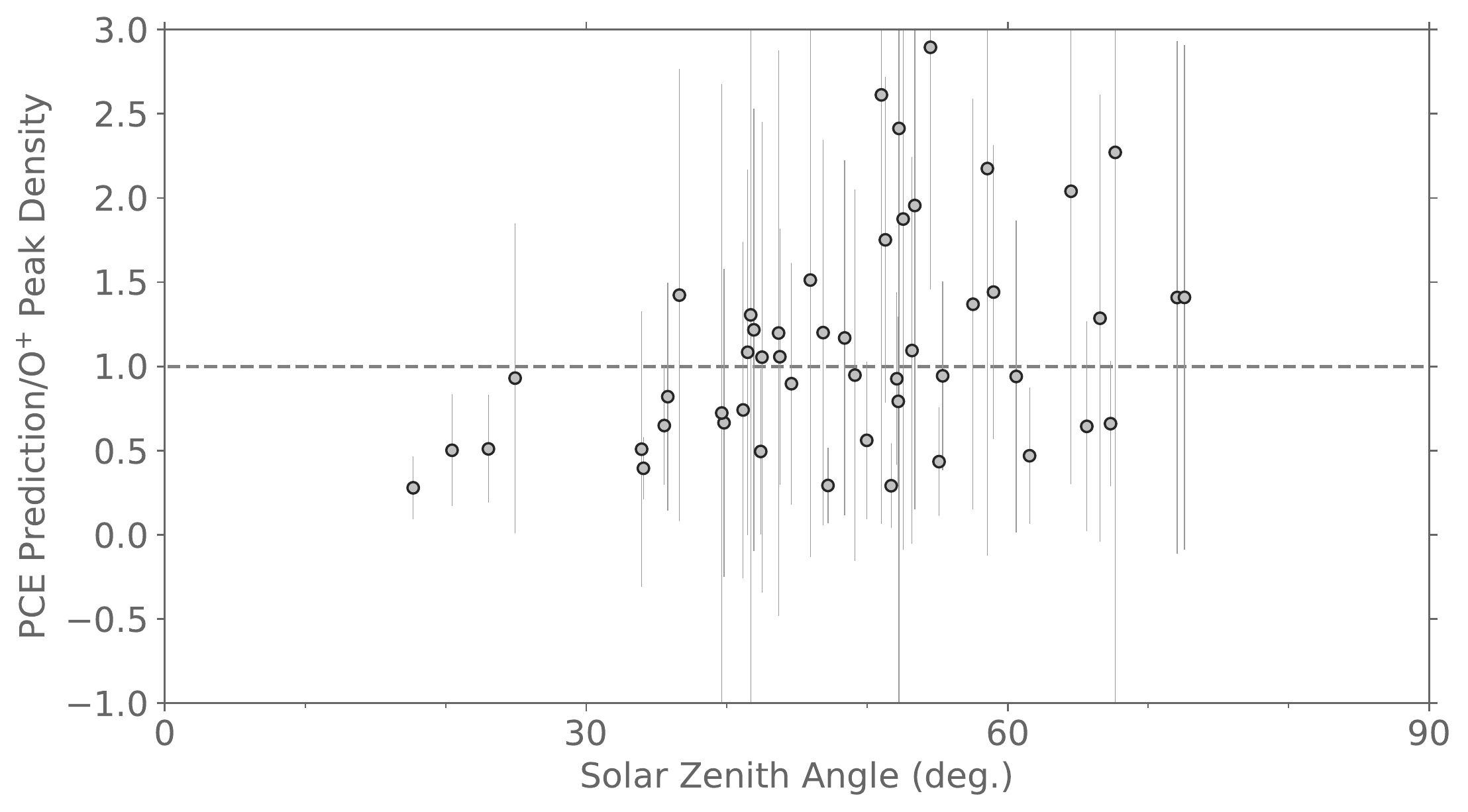}  
\caption{A comparison between the observed O$^+$ peak density and the photochemical equilibrium (PCE) prediction of the O$^+$ peak density as a function of SZA. The dashed line at 1.0 marks where the PCE prediction would exactly match the observed value.}
\label{pcepce}
\end{figure}

We now use this equation to test the extent to which the peak O$^+$ density is consistent with the PCE prediction. Figure~\ref{pcepce} shows a comparison (ratios) between the observed O$^+$ peak density and the O$^+$ peak density predicted by Eq.~\ref{pce} for SZA$<$80$^{\circ}$. For the PCE calculations, we used the derived average values of $J_0$, $[\mathrm{O}]$, $[\mathrm{CO}_{2}]$, and O$^+$ where all quantities were taken at the O$^+$ peak. The majority of predicted values are accurate within a factor of 2.5 and most of the predicted values are within 1-$\sigma$ of 1.0. This suggests good agreement between the measured values and the PCE prediction and implies that PCE is satisfied or nearly satisfied at the O$^+$ peak. The implications of this result will be discussed in Sec.~\ref{sec:discussions}.

\subsection{The Topside O$^+$/O$_2^+$ Ratio}
\label{sec:ratio}

	We now switch from focusing on the O$^+$ peak to focusing on the composition of the topside ionosphere. Our aim is to quantify the O$^+$/O$_2^+$ ratio at altitudes above the O$^+$ peak and to determine how the ratio varies with SZA and season.
    
\begin{figure}[ht!]
 \centering
\includegraphics[width=1.0\textwidth]{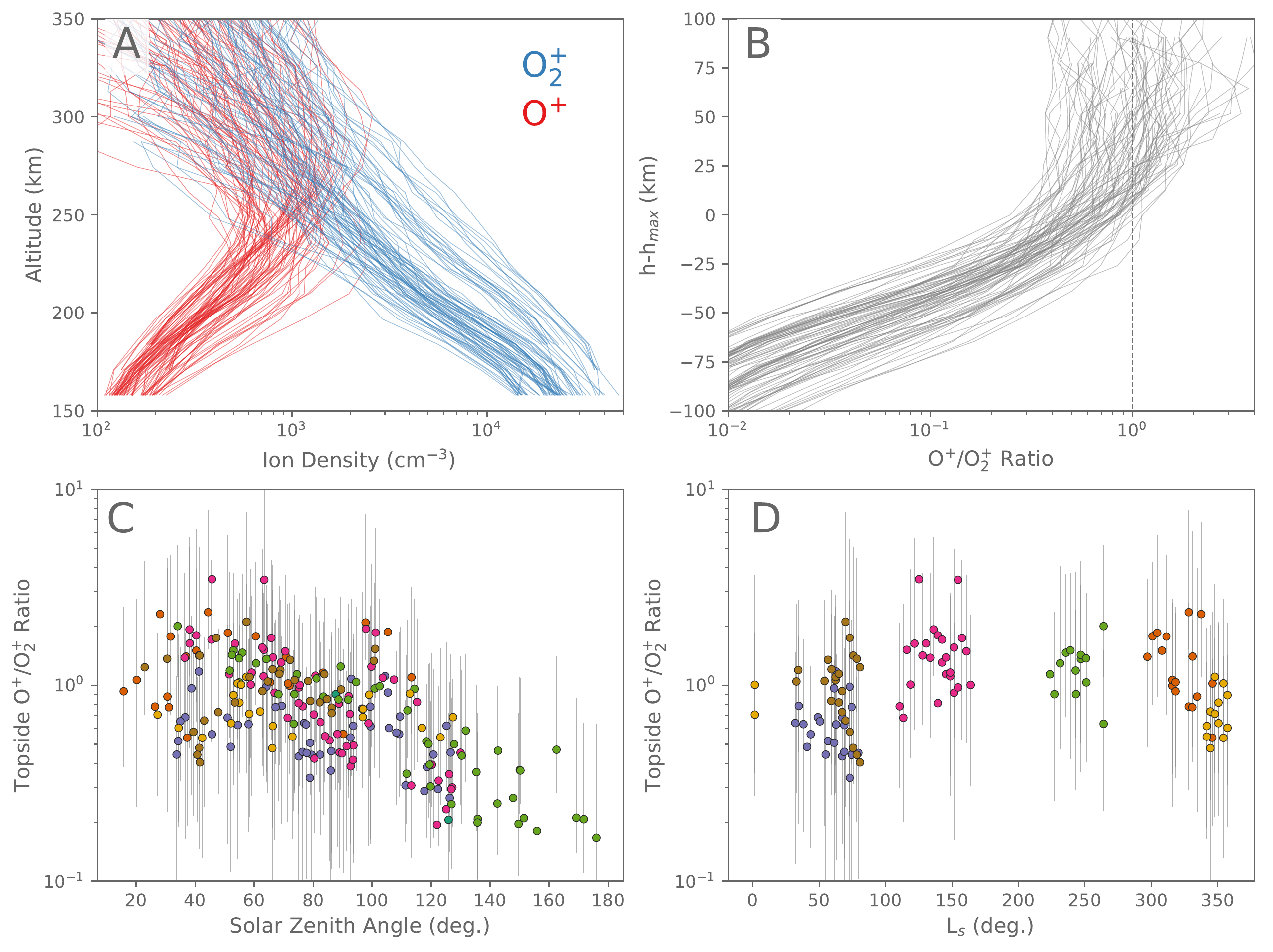}  
\caption{\textbf{A)} Average dayside O$^+$ and O$_2^+$ profiles at SZA$<$80$^{\circ}$. \textbf{B)} The corresponding O$^+$/O$_2^+$ ratios plotted as a function of altitude relative to the O$^+$ peak. The dashed line shows a ratio of 1.0. {C)} The topside O$^+$/O$_2^+$ ratio as a function of SZA. \textbf{D)} The topside O$^+$/O$_2^+$ ratio as a function of L$_s$ for data with SZA$<$80$^{\circ}$. The colors in Panels C and D group data from similar time periods and are the same as a in Fig.~\ref{nmseason}.}
\label{ratio}
\end{figure}

    Figure~\ref{ratio}A shows the average O$^+$ and O$_2^+$ profiles for SZA$<$80$^{\circ}$. Above the O$^+$ peak the O$^+$ and O$_2^+$ densities decrease exponentially with altitude, usually with a common scale height due to the large vertical gradient in the plasma temperature at high altitudes \citep{benna2015a,fox2015}. Here we consider the topside ionosphere to be this region above the O$^+$ peak.
	
	 Figure~\ref{ratio}B shows the corresponding O$^+$/O$_2^+$ ratios. Instead of plotting the ratios against altitude, they are plotted against $h-h_{\mathrm{max}}$, where $h$ is altitude and h$_{\mathrm{max}}$ is the altitude of the O$^+$ peak. The O$^+$/O$_2^+$ ratio increases with altitude until reaching a constant value $\sim$25 km above the O$^+$ peak. We will refer to the O$^+$/O$_2^+$ ratio in this constant region 25 km above the O$^+$ peak as the ``topside ratio''.

	The value of the topside ratio varies significantly from profile to profile having a mean value of 1.1 $\pm$ 0.6. Figure~\ref{ratio}C shows that the topside O$^+$/O$_2^+$ ratio decreases with increasing SZA from $\sim$1.0 at SZA=60$^{\circ}$ to $\sim$0.30 at SZA$>$120$^{\circ}$. Figure~\ref{ratio}D shows the topside ratio as a function of L$_s$ for data with SZA$<$80$^{\circ}$. The topside ratio is somewhat larger between L$_s$=100$^{\circ}$-320$^{\circ}$ when Mars is closest to the Sun but this variation is small given the size of the error bars. The most striking feature in Figs.~\ref{ratio}C and D are the size of the error bars, which represent the  25\% and 75\% quartiles of the ratio over $\sim$15 periapsis passes. The large error bars indicate a highly variable ion composition in the topside ionosphere that changes on timescale of days or less.

\section{Discussion}
\label{sec:discussions}

\subsection{Comparisons Between the O$^+$ Peak and the O$_2^+$ Peak}
\label{sec:M2}
	The O$^+$ peak varies quite differently than the main peak (M2 peak) of the ionosphere, which includes both the electron density peak and the O$_2^+$ peak. At the main peak, the M2 peak electron density and peak altitude vary with SZA and EUV flux as predicted by Chapman theory for a simple photochemical layer:
	\begin{equation}
	\label{nm2}
	 N_{\mathrm{max},M2} \propto \sqrt{F_{0} \cos{\mathrm{(SZA)}}}
	\end{equation}
	\begin{equation}
	\label{hm2}
	 h_{\mathrm{max},M2} \propto H\ln{\left[\frac{1}{\cos{(\mathrm{SZA})}}\right]}
	\end{equation}
	
\noindent{where} $N_{\mathrm{max},M2}$ is the M2 peak density, $F_{0}$ is the ionizing EUV flux at the top of the atmosphere, $h_{\mathrm{max},M2}$ is the M2 peak altitude \citep{schunk2009}, and $H$ is the neutral CO$_2$ scale height. Equation~\ref{nm2} predicts that the M2 peak density increases with the square-root of the EUV flux and decreases with increasing SZA. Equation~\ref{hm2} predicts that the M2 peak altitude increases with increasing SZA. Tests of these predictions have found that the M2 peak density and peak altitude largely satisfy Eqs.~\ref{nm2} and \ref{hm2} \citep{morgan2008,withers2009a,girazian2013,mendillo2013,fallows2015,vogt2017}. 
        
  The simplifying assumptions needed to derive Eqs.~\ref{nm2} and \ref{hm2} include that the M2 peak is in PCE (photochemical equilibrium) and composed of a single molecular ion species (O$_2^+$) that is destroyed entirely by DR (dissociative recombination) with an electron (O$_2^+$ + e$^{-} \rightarrow$ O + O). Given these assumptions, the M2 peak forms at the altitude where the slant optical depth for ionizing EUV photons is equal to one \citep{withers2009a}. The $\cos{\mathrm{(SZA)}}$ terms in Eqs.~\ref{nm2} and \ref{hm2} arise from this rule; it is the main reason why the M2 peak density and peak altitude vary with SZA.
	
	The $\cos{\mathrm{(SZA)}}$ term, however, is not predicted for O$^+$ under PCE conditions (Eq.~\ref{pce}). The main reason being that O$^+$ is an atomic ion (not molecular) so it is mainly destroyed by reacting with neutral molecules instead of DR (e.g., Eq.~\ref{lrate}). Consequently, the O$^+$ peak forms at a higher altitude than the M2 peak, where the atmosphere is optically thin to EUV photons. Our results confirm this: the peak O$^+$ density has no strong dependence on SZA up to SZA=90$^{\circ}$ (Fig.~\ref{nmax}), a much weaker trend than the M2 peak. Instead, the O$^+$ peak density variations are controlled primarily by the O/CO$_2$ ratio at the O$^+$ peak, just as Eq.~\ref{pce} predicts. We confirmed this with our analysis in Sec.~\ref{peakdensity}.
    
  Furthermore, the M2 peak altitude increases near the the terminator to satisfy the condition that the optical depth of EUV photons is equal to one (Eq.~\ref{hm2}). By contrast, the O$^+$ peak altitude does the opposite, it decreases near the terminator, perhaps in order to track a level of roughly constant neutral pressure, or due to the increased role of transport driven by day-night pressure gradients. In any case, because the O$^+$ peak is not formed at unit optical depth, it varies differently with SZA than the M2 peak.

\subsection{Comparisons Between the O$^+$ Peak and the Terrestrial F2 Peak}
\label{sec:F2}
    
    It is interesting to consider similarities between the O$^+$ peak at Mars and the F2 peak at Earth, which is also composed of O$^+$ ions. In both cases PCE predicts that the O$^+$ density increases indefinitely with altitude. At Mars this occurs because the neutral O density decreases with altitude at a slower rate than the CO$_2$ density. Consequently, the O$^+$ photoionization rate increases with altitude at a faster rate than the O$^+$ chemical loss rate (Eq.~\ref{pce}). At Earth, a similar situation occurs except that O$^+$ is destroyed by reacting with N$_2$ and O$_2$ rather than CO$_2$ \citep{schunk2009}.
    
    Under strict PCE conditions, then, an O$^+$ peak cannot form at either planet \citep{rishbeth1963,fox2001,schunk2009,mendillo2011,chaufray2014}. The formation of the O$^+$ peak requires transport processes such as vertical diffusion, which has been confirmed with numerical models of the ionosphere \citep{rishbeth1963,mendillo2011,chaufray2014}. 
   
    At both planets, the O$^+$ density increases with altitude in the PCE region, forms a peak near the PCE boundary, then decreases with altitude above the peak where transport processes dominate \citep{rishbeth1963,schunk2009}. As our results show, the O$^+$ peak density is still consistent with the the predictions from PCE theory (Eq.~\ref{pce}), which is also true for the terrestrial F2 layer \citep{rishbeth1963}.  
   
   There is another interesting similarity between the O$^+$ layers at Earth and Mars: both peaks form at a roughly constant neutral atmospheric pressure level. As shown here for Mars, the dayside O$^+$ peak forms at a fixed pressure level of 10$^{(-8.7 \pm 0.4)}$ Pa, while at Earth, the O$^+$ peak forms at a fixed pressure level of 3 $\times$ 10$^{-5}$ Pa \citep{rishbeth1989}. 
    

	
    
\section{Conclusions}
 \label{conclusions}
	In conclusion, the seasonal, SZA, and solar flux variations of the O$^+$ peak density, O$^+$ peak altitude, and topside O$^+$/O$_2^+$ ratio have -- for the first -- time been constrained by observations. The major findings include that the O$^+$ peak forms at a roughly constant pressure level, the O$^+$ peak density and peak altitude vary differently with SZA than the main peak of the ionosphere, and the dayside O$^+$/O$_2^+$ ratio in the topside ionosphere is highly variable but, on average, is close to one. These results can be used to validate numerical models of the ionosphere, of which many exist. Additional analyses of MAVEN data and numerical simulations will also be beneficial for determining how the O$^+$ variations presented here affect ion escape at Mars.

\acknowledgments
Z.G. thanks Andrew Nagy, Paul Withers, and Joseph Grebowsky for useful suggestions regarding this work. Z.G.'s research was supported by an appointment to the NASA Postdoctoral Program at the NASA Goddard Space Flight Center, administered by Universities Space Research Association under contract with NASA. This research was also supported by the MAVEN mission through NASA headquarters. The data used in this publication are publicly available and can be downloaded from the MAVEN Science Data Center (\url{https://lasp.colorado.edu/maven/sdc/public/}).

%
%
%
%
%
%
%
%
%


\begin{thebibliography}{}

\bibitem [\protect \citeauthoryear {%
{Benna}%
\ \protect \BOthers {.}}{%
{Benna}%
\ \protect \BOthers {.}}{%
{\protect \APACyear {2015}}%
}]{%
benna2015a}
\APACinsertmetastar {%
benna2015a}%
\begin{APACrefauthors}%
{Benna}, M.%
, {Mahaffy}, P\BPBI R.%
, {Grebowsky}, J\BPBI M.%
, {Fox}, J\BPBI L.%
, {Yelle}, R\BPBI V.%
\BCBL {}\ \BBA {} {Jakosky}, B\BPBI M.%
\end{APACrefauthors}%
\unskip\
\newblock
\APACrefYearMonthDay{2015}{}{}.
\newblock
{\BBOQ}\APACrefatitle {{First measurements of composition and dynamics of the
  {M}artian ionosphere by {MAVEN}'s {N}eutral {G}as and {I}on {M}ass
  {S}pectrometer}} {{First measurements of composition and dynamics of the
  {M}artian ionosphere by {MAVEN}'s {N}eutral {G}as and {I}on {M}ass
  {S}pectrometer}}.{\BBCQ}
\newblock
\APACjournalVolNumPages{Geophys. Res. Lett.}{42}{}{8958-8965}.
\newblock
\begin{APACrefDOI} \doi{10.1002/2015GL066146} \end{APACrefDOI}
\PrintBackRefs{\CurrentBib}

\bibitem [\protect \citeauthoryear {%
{Bougher}%
, {Engel}%
, {Roble}%
\BCBL {}\ \BBA {} {Foster}%
}{%
{Bougher}%
\ \protect \BOthers {.}}{%
{\protect \APACyear {2000}}%
}]{%
bougher2000a}
\APACinsertmetastar {%
bougher2000a}%
\begin{APACrefauthors}%
{Bougher}, S\BPBI W.%
, {Engel}, S.%
, {Roble}, R\BPBI G.%
\BCBL {}\ \BBA {} {Foster}, B.%
\end{APACrefauthors}%
\unskip\
\newblock
\APACrefYearMonthDay{2000}{}{}.
\newblock
{\BBOQ}\APACrefatitle {{Comparative terrestrial planet thermospheres 3. {S}olar
  cycle variation of global structure and winds at solstices}} {{Comparative
  terrestrial planet thermospheres 3. {S}olar cycle variation of global
  structure and winds at solstices}}.{\BBCQ}
\newblock
\APACjournalVolNumPages{J. Geophys. Res.}{105}{}{17669-17692}.
\newblock
\begin{APACrefDOI} \doi{10.1029/1999JE001232} \end{APACrefDOI}
\PrintBackRefs{\CurrentBib}

\bibitem [\protect \citeauthoryear {%
{Bougher}%
\ \protect \BOthers {.}}{%
{Bougher}%
\ \protect \BOthers {.}}{%
{\protect \APACyear {2015}}%
}]{%
bougher2015}
\APACinsertmetastar {%
bougher2015}%
\begin{APACrefauthors}%
{Bougher}, S\BPBI W.%
, {Pawlowski}, D.%
, {Bell}, J\BPBI M.%
, {Nelli}, S.%
, {McDunn}, T.%
, {Murphy}, J\BPBI R.%
\BDBL {}{Ridley}, A.%
\end{APACrefauthors}%
\unskip\
\newblock
\APACrefYearMonthDay{2015}{}{}.
\newblock
{\BBOQ}\APACrefatitle {{Mars {G}lobal {I}onosphere-{T}hermosphere {M}odel:
  {S}olar cycle, seasonal, and diurnal variations of the {M}ars upper
  atmosphere}} {{Mars {G}lobal {I}onosphere-{T}hermosphere {M}odel: {S}olar
  cycle, seasonal, and diurnal variations of the {M}ars upper
  atmosphere}}.{\BBCQ}
\newblock
\APACjournalVolNumPages{J. Geophys. Res.}{120}{}{311-342}.
\newblock
\begin{APACrefDOI} \doi{10.1002/2014JE004715} \end{APACrefDOI}
\PrintBackRefs{\CurrentBib}

\bibitem [\protect \citeauthoryear {%
Bougher%
\ \protect \BOthers {.}}{%
Bougher%
\ \protect \BOthers {.}}{%
{\protect \APACyear {2017}}%
}]{%
bougher2017}
\APACinsertmetastar {%
bougher2017}%
\begin{APACrefauthors}%
Bougher, S\BPBI W.%
, Roeten, K\BPBI J.%
, Olsen, K.%
, Mahaffy, P\BPBI R.%
, Benna, M.%
, Elrod, M.%
\BDBL {}Jakosky, B\BPBI M.%
\end{APACrefauthors}%
\unskip\
\newblock
\APACrefYearMonthDay{2017}{}{}.
\newblock
{\BBOQ}\APACrefatitle {The structure and variability of {M}ars dayside
  thermosphere from {MAVEN NGIMS} and {IUVS} measurements: Seasonal and solar
  activity trends in scale heights and temperatures} {The structure and
  variability of {M}ars dayside thermosphere from {MAVEN NGIMS} and {IUVS}
  measurements: Seasonal and solar activity trends in scale heights and
  temperatures}.{\BBCQ}
\newblock
\APACjournalVolNumPages{J. Geophys. Res.}{122}{}{1296--1313}.
\newblock
\begin{APACrefDOI} \doi{10.1002/2016JA023454} \end{APACrefDOI}
\PrintBackRefs{\CurrentBib}

\bibitem [\protect \citeauthoryear {%
{Brain}%
, {Bagenal}%
, {Ma}%
, {Nilsson}%
\BCBL {}\ \BBA {} {Stenberg Wieser}%
}{%
{Brain}%
\ \protect \BOthers {.}}{%
{\protect \APACyear {2016}}%
}]{%
brain2016}
\APACinsertmetastar {%
brain2016}%
\begin{APACrefauthors}%
{Brain}, D\BPBI A.%
, {Bagenal}, F.%
, {Ma}, Y\BHBI J.%
, {Nilsson}, H.%
\BCBL {}\ \BBA {} {Stenberg Wieser}, G.%
\end{APACrefauthors}%
\unskip\
\newblock
\APACrefYearMonthDay{2016}{}{}.
\newblock
{\BBOQ}\APACrefatitle {{Atmospheric escape from unmagnetized bodies}}
  {{Atmospheric escape from unmagnetized bodies}}.{\BBCQ}
\newblock
\APACjournalVolNumPages{J. Geophys. Res.}{121}{}{2364-2385}.
\newblock
\begin{APACrefDOI} \doi{10.1002/2016JE005162} \end{APACrefDOI}
\PrintBackRefs{\CurrentBib}

\bibitem [\protect \citeauthoryear {%
{Carlsson}%
\ \protect \BOthers {.}}{%
{Carlsson}%
\ \protect \BOthers {.}}{%
{\protect \APACyear {2006}}%
}]{%
carlsson2006}
\APACinsertmetastar {%
carlsson2006}%
\begin{APACrefauthors}%
{Carlsson}, E.%
, {Fedorov}, A.%
, {Barabash}, S.%
, {Budnik}, E.%
, {Grigoriev}, A.%
, {Gunell}, H.%
\BDBL {}{Dierker}, C.%
\end{APACrefauthors}%
\unskip\
\newblock
\APACrefYearMonthDay{2006}{}{}.
\newblock
{\BBOQ}\APACrefatitle {{Mass composition of the escaping plasma at {M}ars}}
  {{Mass composition of the escaping plasma at {M}ars}}.{\BBCQ}
\newblock
\APACjournalVolNumPages{Icarus}{182}{}{320-328}.
\newblock
\begin{APACrefDOI} \doi{10.1016/j.icarus.2005.09.020} \end{APACrefDOI}
\PrintBackRefs{\CurrentBib}

\bibitem [\protect \citeauthoryear {%
{Chaufray}%
\ \protect \BOthers {.}}{%
{Chaufray}%
\ \protect \BOthers {.}}{%
{\protect \APACyear {2014}}%
}]{%
chaufray2014}
\APACinsertmetastar {%
chaufray2014}%
\begin{APACrefauthors}%
{Chaufray}, J\BHBI Y.%
, {Gonzalez-Galindo}, F.%
, {Forget}, F.%
, {Lopez-Valverde}, M.%
, {Leblanc}, F.%
, {Modolo}, R.%
\BDBL {}{Witasse}, O.%
\end{APACrefauthors}%
\unskip\
\newblock
\APACrefYearMonthDay{2014}{}{}.
\newblock
{\BBOQ}\APACrefatitle {{Three-dimensional {M}artian ionosphere model: {II}.
  {E}ffect of transport processes due to pressure gradients}}
  {{Three-dimensional {M}artian ionosphere model: {II}. {E}ffect of transport
  processes due to pressure gradients}}.{\BBCQ}
\newblock
\APACjournalVolNumPages{J. Geophys. Res.}{119}{}{1614-1636}.
\newblock
\begin{APACrefDOI} \doi{10.1002/2013JE004551} \end{APACrefDOI}
\PrintBackRefs{\CurrentBib}

\bibitem [\protect \citeauthoryear {%
{Dong}%
\ \protect \BOthers {.}}{%
{Dong}%
\ \protect \BOthers {.}}{%
{\protect \APACyear {2017}}%
}]{%
dong2017}
\APACinsertmetastar {%
dong2017}%
\begin{APACrefauthors}%
{Dong}, Y.%
, {Fang}, X.%
, {Brain}, D\BPBI A.%
, {McFadden}, J\BPBI P.%
, {Halekas}, J\BPBI S.%
, {Connerney}, J\BPBI E\BPBI P.%
\BDBL {}{Jakosky}, B\BPBI M.%
\end{APACrefauthors}%
\unskip\
\newblock
\APACrefYearMonthDay{2017}{}{}.
\newblock
{\BBOQ}\APACrefatitle {{Seasonal variability of {M}artian ion escape through
  the plume and tail from {MAVEN} observations}} {{Seasonal variability of
  {M}artian ion escape through the plume and tail from {MAVEN}
  observations}}.{\BBCQ}
\newblock
\APACjournalVolNumPages{J. Geophys. Res.}{122}{}{4009-4022}.
\newblock
\begin{APACrefDOI} \doi{10.1002/2016JA023517} \end{APACrefDOI}
\PrintBackRefs{\CurrentBib}

\bibitem [\protect \citeauthoryear {%
Dubinin%
\ \protect \BOthers {.}}{%
Dubinin%
\ \protect \BOthers {.}}{%
{\protect \APACyear {2018}}%
}]{%
dubinin2018}
\APACinsertmetastar {%
dubinin2018}%
\begin{APACrefauthors}%
Dubinin, E.%
, Fraenz, M.%
, Pätzold, M.%
, McFadden, J.%
, Halekas, J.%
, Connerney, J.%
\BDBL {}Zelenyi, L.%
\end{APACrefauthors}%
\unskip\
\newblock
\APACrefYearMonthDay{2018}{}{}.
\newblock
{\BBOQ}\APACrefatitle {Martian ionosphere observed by {MAVEN}. 3. {I}nfluence
  of solar wind and {IMF} on upper ionosphere} {Martian ionosphere observed by
  {MAVEN}. 3. {I}nfluence of solar wind and {IMF} on upper ionosphere}.{\BBCQ}
\newblock
\APACjournalVolNumPages{Planet. Space Sci.}{}{}{-}.
\newblock
\begin{APACrefURL}
  \url{https://www.sciencedirect.com/science/article/pii/S0032063317304361}
  \end{APACrefURL}
\newblock
\begin{APACrefDOI} \doi{https://doi.org/10.1016/j.pss.2018.03.016}
  \end{APACrefDOI}
\PrintBackRefs{\CurrentBib}

\bibitem [\protect \citeauthoryear {%
{Elrod}%
\ \protect \BOthers {.}}{%
{Elrod}%
\ \protect \BOthers {.}}{%
{\protect \APACyear {2017}}%
}]{%
elrod2017}
\APACinsertmetastar {%
elrod2017}%
\begin{APACrefauthors}%
{Elrod}, M\BPBI K.%
, {Bougher}, S.%
, {Bell}, J.%
, {Mahaffy}, P\BPBI R.%
, {Benna}, M.%
, {Stone}, S.%
\BDBL {}{Jakosky}, B.%
\end{APACrefauthors}%
\unskip\
\newblock
\APACrefYearMonthDay{2017}{}{}.
\newblock
{\BBOQ}\APACrefatitle {{He bulge revealed: {H}e and {CO}$_{2}$ diurnal and
  seasonal variations in the upper atmosphere of Mars as detected by {MAVEN
  NGIM}}} {{He bulge revealed: {H}e and {CO}$_{2}$ diurnal and seasonal
  variations in the upper atmosphere of Mars as detected by {MAVEN
  NGIM}}}.{\BBCQ}
\newblock
\APACjournalVolNumPages{J. Geophys. Res.}{122}{}{2564-2573}.
\newblock
\begin{APACrefDOI} \doi{10.1002/2016JA023482} \end{APACrefDOI}
\PrintBackRefs{\CurrentBib}

\bibitem [\protect \citeauthoryear {%
{Eparvier}%
, {Chamberlin}%
, {Woods}%
\BCBL {}\ \BBA {} {Thiemann}%
}{%
{Eparvier}%
\ \protect \BOthers {.}}{%
{\protect \APACyear {2015}}%
}]{%
eparvier2015}
\APACinsertmetastar {%
eparvier2015}%
\begin{APACrefauthors}%
{Eparvier}, F\BPBI G.%
, {Chamberlin}, P\BPBI C.%
, {Woods}, T\BPBI N.%
\BCBL {}\ \BBA {} {Thiemann}, E\BPBI M\BPBI B.%
\end{APACrefauthors}%
\unskip\
\newblock
\APACrefYearMonthDay{2015}{}{}.
\newblock
{\BBOQ}\APACrefatitle {{The {S}olar {E}xtreme {U}ltraviolet {M}onitor for
  {MAVEN}}} {{The {S}olar {E}xtreme {U}ltraviolet {M}onitor for
  {MAVEN}}}.{\BBCQ}
\newblock
\APACjournalVolNumPages{Space Sci. Rev.}{195}{}{293-301}.
\newblock
\begin{APACrefDOI} \doi{10.1007/s11214-015-0195-2} \end{APACrefDOI}
\PrintBackRefs{\CurrentBib}

\bibitem [\protect \citeauthoryear {%
{Fallows}%
, {Withers}%
\BCBL {}\ \BBA {} {Matta}%
}{%
{Fallows}%
\ \protect \BOthers {.}}{%
{\protect \APACyear {2015}}%
}]{%
fallows2015}
\APACinsertmetastar {%
fallows2015}%
\begin{APACrefauthors}%
{Fallows}, K.%
, {Withers}, P.%
\BCBL {}\ \BBA {} {Matta}, M.%
\end{APACrefauthors}%
\unskip\
\newblock
\APACrefYearMonthDay{2015}{}{}.
\newblock
{\BBOQ}\APACrefatitle {{An observational study of the influence of solar zenith
  angle on properties of the {M}1 layer of the {M}ars ionosphere}} {{An
  observational study of the influence of solar zenith angle on properties of
  the {M}1 layer of the {M}ars ionosphere}}.{\BBCQ}
\newblock
\APACjournalVolNumPages{J. Geophys. Res.}{120}{}{1299-1310}.
\newblock
\begin{APACrefDOI} \doi{10.1002/2014JA020750} \end{APACrefDOI}
\PrintBackRefs{\CurrentBib}

\bibitem [\protect \citeauthoryear {%
{Fox}%
}{%
{Fox}%
}{%
{\protect \APACyear {2004}}%
}]{%
fox2004}
\APACinsertmetastar {%
fox2004}%
\begin{APACrefauthors}%
{Fox}, J\BPBI L.%
\end{APACrefauthors}%
\unskip\
\newblock
\APACrefYearMonthDay{2004}{}{}.
\newblock
{\BBOQ}\APACrefatitle {Response of the Martian thermosphere/ionosphere to
  enhanced fluxes of solar soft X rays} {Response of the martian
  thermosphere/ionosphere to enhanced fluxes of solar soft x rays}.{\BBCQ}
\newblock
\APACjournalVolNumPages{J. Geophys. Res.}{109}{}{11310}.
\newblock
\begin{APACrefDOI} \doi{10.1029/2004JA010380} \end{APACrefDOI}
\PrintBackRefs{\CurrentBib}

\bibitem [\protect \citeauthoryear {%
{Fox}%
}{%
{Fox}%
}{%
{\protect \APACyear {2009}}%
}]{%
fox2009a}
\APACinsertmetastar {%
fox2009a}%
\begin{APACrefauthors}%
{Fox}, J\BPBI L.%
\end{APACrefauthors}%
\unskip\
\newblock
\APACrefYearMonthDay{2009}{}{}.
\newblock
{\BBOQ}\APACrefatitle {{Morphology of the dayside ionosphere of {M}ars:
  {I}mplications for ion outflows}} {{Morphology of the dayside ionosphere of
  {M}ars: {I}mplications for ion outflows}}.{\BBCQ}
\newblock
\APACjournalVolNumPages{J. Geophys. Res.}{114}{}{12005}.
\newblock
\begin{APACrefDOI} \doi{10.1029/2009JE003432} \end{APACrefDOI}
\PrintBackRefs{\CurrentBib}

\bibitem [\protect \citeauthoryear {%
{Fox}%
}{%
{Fox}%
}{%
{\protect \APACyear {2015}}%
}]{%
fox2015}
\APACinsertmetastar {%
fox2015}%
\begin{APACrefauthors}%
{Fox}, J\BPBI L.%
\end{APACrefauthors}%
\unskip\
\newblock
\APACrefYearMonthDay{2015}{}{}.
\newblock
{\BBOQ}\APACrefatitle {{The chemistry of protonated species in the martian
  ionosphere}} {{The chemistry of protonated species in the martian
  ionosphere}}.{\BBCQ}
\newblock
\APACjournalVolNumPages{Icarus}{252}{}{366-392}.
\newblock
\begin{APACrefDOI} \doi{10.1016/j.icarus.2015.01.010} \end{APACrefDOI}
\PrintBackRefs{\CurrentBib}

\bibitem [\protect \citeauthoryear {%
{Fox}%
\ \BBA {} {Sung}%
}{%
{Fox}%
\ \BBA {} {Sung}%
}{%
{\protect \APACyear {2001}}%
}]{%
fox2001}
\APACinsertmetastar {%
fox2001}%
\begin{APACrefauthors}%
{Fox}, J\BPBI L.%
\BCBT {}\ \BBA {} {Sung}, K\BPBI Y.%
\end{APACrefauthors}%
\unskip\
\newblock
\APACrefYearMonthDay{2001}{{\APACmonth{10}}}{}.
\newblock
{\BBOQ}\APACrefatitle {Solar activity variations of the {V}enus
  thermosphere/ionosphere} {Solar activity variations of the {V}enus
  thermosphere/ionosphere}.{\BBCQ}
\newblock
\APACjournalVolNumPages{J. Geophys. Res.}{106}{}{21305-21336}.
\newblock
\begin{APACrefDOI} \doi{10.1029/2001JA000069} \end{APACrefDOI}
\PrintBackRefs{\CurrentBib}

\bibitem [\protect \citeauthoryear {%
Fox%
, Zhou%
\BCBL {}\ \BBA {} Bougher%
}{%
Fox%
\ \protect \BOthers {.}}{%
{\protect \APACyear {1996}}%
}]{%
fox1996}
\APACinsertmetastar {%
fox1996}%
\begin{APACrefauthors}%
Fox, J\BPBI L.%
, Zhou, P.%
\BCBL {}\ \BBA {} Bougher, S\BPBI W.%
\end{APACrefauthors}%
\unskip\
\newblock
\APACrefYearMonthDay{1996}{}{}.
\newblock
{\BBOQ}\APACrefatitle {The martian thermosphere/ionosphere at high and low
  solar activities} {The martian thermosphere/ionosphere at high and low solar
  activities}.{\BBCQ}
\newblock
\APACjournalVolNumPages{Adv. Space Res.}{17}{}{203-218}.
\newblock
\begin{APACrefDOI} \doi{10.1016/0273-1177(95)00751-Y} \end{APACrefDOI}
\PrintBackRefs{\CurrentBib}

\bibitem [\protect \citeauthoryear {%
{Girazian}%
\ \protect \BOthers {.}}{%
{Girazian}%
\ \protect \BOthers {.}}{%
{\protect \APACyear {2017a}}%
}]{%
girazian2017a}
\APACinsertmetastar {%
girazian2017a}%
\begin{APACrefauthors}%
{Girazian}, Z.%
, {Mahaffy}, P.%
, {Lillis}, R\BPBI J.%
, {Benna}, M.%
, {Elrod}, M.%
, {Fowler}, C\BPBI M.%
\BCBL {}\ \BBA {} {Mitchell}, D\BPBI L.%
\end{APACrefauthors}%
\unskip\
\newblock
\APACrefYearMonthDay{2017a}{}{}.
\newblock
{\BBOQ}\APACrefatitle {{Ion {D}ensities in the {N}ightside {I}onosphere of
  {M}ars: {E}ffects of {E}lectron {I}mpact {I}onization}} {{Ion {D}ensities in
  the {N}ightside {I}onosphere of {M}ars: {E}ffects of {E}lectron {I}mpact
  {I}onization}}.{\BBCQ}
\newblock
\APACjournalVolNumPages{Geophys. Res. Lett.}{44}{}{11}.
\newblock
\begin{APACrefDOI} \doi{10.1002/2017GL075431} \end{APACrefDOI}
\PrintBackRefs{\CurrentBib}

\bibitem [\protect \citeauthoryear {%
{Girazian}%
\ \protect \BOthers {.}}{%
{Girazian}%
\ \protect \BOthers {.}}{%
{\protect \APACyear {2017b}}%
}]{%
girazian2017}
\APACinsertmetastar {%
girazian2017}%
\begin{APACrefauthors}%
{Girazian}, Z.%
, {Mahaffy}, P\BPBI R.%
, {Lillis}, R\BPBI J.%
, {Benna}, M.%
, {Elrod}, M.%
\BCBL {}\ \BBA {} {Jakosky}, B\BPBI M.%
\end{APACrefauthors}%
\unskip\
\newblock
\APACrefYearMonthDay{2017b}{}{}.
\newblock
{\BBOQ}\APACrefatitle {{Nightside ionosphere of {M}ars: {C}omposition, vertical
  structure, and variability}} {{Nightside ionosphere of {M}ars: {C}omposition,
  vertical structure, and variability}}.{\BBCQ}
\newblock
\APACjournalVolNumPages{J. Geophys. Res.}{122}{}{4712-4725}.
\newblock
\begin{APACrefDOI} \doi{10.1002/2016JA023508} \end{APACrefDOI}
\PrintBackRefs{\CurrentBib}

\bibitem [\protect \citeauthoryear {%
{Girazian}%
\ \BBA {} {Withers}%
}{%
{Girazian}%
\ \BBA {} {Withers}%
}{%
{\protect \APACyear {2013}}%
}]{%
girazian2013}
\APACinsertmetastar {%
girazian2013}%
\begin{APACrefauthors}%
{Girazian}, Z.%
\BCBT {}\ \BBA {} {Withers}, P.%
\end{APACrefauthors}%
\unskip\
\newblock
\APACrefYearMonthDay{2013}{}{}.
\newblock
{\BBOQ}\APACrefatitle {The dependence of peak electron density in the
  ionosphere of {M}ars on solar irradiance} {The dependence of peak electron
  density in the ionosphere of {M}ars on solar irradiance}.{\BBCQ}
\newblock
\APACjournalVolNumPages{Geophys. Res. Lett.}{40}{}{1960-1964}.
\newblock
\begin{APACrefDOI} \doi{10.1002/grl.50344} \end{APACrefDOI}
\PrintBackRefs{\CurrentBib}

\bibitem [\protect \citeauthoryear {%
{Girazian}%
\ \BBA {} {Withers}%
}{%
{Girazian}%
\ \BBA {} {Withers}%
}{%
{\protect \APACyear {2015}}%
}]{%
girazian2015b}
\APACinsertmetastar {%
girazian2015b}%
\begin{APACrefauthors}%
{Girazian}, Z.%
\BCBT {}\ \BBA {} {Withers}, P.%
\end{APACrefauthors}%
\unskip\
\newblock
\APACrefYearMonthDay{2015}{}{}.
\newblock
{\BBOQ}\APACrefatitle {{An empirical model of the extreme ultraviolet solar
  spectrum as a function of {F}10.7}} {{An empirical model of the extreme
  ultraviolet solar spectrum as a function of {F}10.7}}.{\BBCQ}
\newblock
\APACjournalVolNumPages{J. Geophys. Res.}{120}{}{6779–6794}.
\newblock
\begin{APACrefDOI} \doi{10.1002/2015JA021436} \end{APACrefDOI}
\PrintBackRefs{\CurrentBib}

\bibitem [\protect \citeauthoryear {%
{Gonz{\'a}lez-Galindo}%
\ \protect \BOthers {.}}{%
{Gonz{\'a}lez-Galindo}%
\ \protect \BOthers {.}}{%
{\protect \APACyear {2013}}%
}]{%
galindo2013}
\APACinsertmetastar {%
galindo2013}%
\begin{APACrefauthors}%
{Gonz{\'a}lez-Galindo}, F.%
, {Chaufray}, J\BHBI Y.%
, {L{\'o}pez-Valverde}, M\BPBI A.%
, {Gilli}, G.%
, {Forget}, F.%
, {Leblanc}, F.%
\BDBL {}{Yagi}, M.%
\end{APACrefauthors}%
\unskip\
\newblock
\APACrefYearMonthDay{2013}{}{}.
\newblock
{\BBOQ}\APACrefatitle {{Three-dimensional {M}artian ionosphere model: I. {T}he
  photochemical ionosphere below 180 km}} {{Three-dimensional {M}artian
  ionosphere model: I. {T}he photochemical ionosphere below 180 km}}.{\BBCQ}
\newblock
\APACjournalVolNumPages{J. Geophys. Res.}{118}{}{2105-2123}.
\newblock
\begin{APACrefDOI} \doi{10.1002/jgre.20150} \end{APACrefDOI}
\PrintBackRefs{\CurrentBib}

\bibitem [\protect \citeauthoryear {%
{Gurnett}%
\ \protect \BOthers {.}}{%
{Gurnett}%
\ \protect \BOthers {.}}{%
{\protect \APACyear {2008}}%
}]{%
gurnett2008}
\APACinsertmetastar {%
gurnett2008}%
\begin{APACrefauthors}%
{Gurnett}, D\BPBI A.%
, {Huff}, R\BPBI L.%
, {Morgan}, D\BPBI D.%
, {Persoon}, A\BPBI M.%
, {Averkamp}, T\BPBI F.%
, {Kirchner}, D\BPBI L.%
\BDBL {}{Picardi}, G.%
\end{APACrefauthors}%
\unskip\
\newblock
\APACrefYearMonthDay{2008}{}{}.
\newblock
{\BBOQ}\APACrefatitle {An overview of radar soundings of the martian ionosphere
  from the {M}ars {E}xpress spacecraft} {An overview of radar soundings of the
  martian ionosphere from the {M}ars {E}xpress spacecraft}.{\BBCQ}
\newblock
\APACjournalVolNumPages{Adv. Space Res.}{41}{}{1335-1346}.
\newblock
\begin{APACrefDOI} \doi{10.1016/j.asr.2007.01.062} \end{APACrefDOI}
\PrintBackRefs{\CurrentBib}

\bibitem [\protect \citeauthoryear {%
{Hanson}%
, {Sanatani}%
\BCBL {}\ \BBA {} {Zuccaro}%
}{%
{Hanson}%
\ \protect \BOthers {.}}{%
{\protect \APACyear {1977}}%
}]{%
hanson1977}
\APACinsertmetastar {%
hanson1977}%
\begin{APACrefauthors}%
{Hanson}, W\BPBI B.%
, {Sanatani}, S.%
\BCBL {}\ \BBA {} {Zuccaro}, D\BPBI R.%
\end{APACrefauthors}%
\unskip\
\newblock
\APACrefYearMonthDay{1977}{}{}.
\newblock
{\BBOQ}\APACrefatitle {The martian ionosphere as observed by the {V}iking
  {R}etarding {P}otential {A}nalyzers} {The martian ionosphere as observed by
  the {V}iking {R}etarding {P}otential {A}nalyzers}.{\BBCQ}
\newblock
\APACjournalVolNumPages{J. Geophys. Res.}{82}{}{4351-4363}.
\newblock
\begin{APACrefDOI} \doi{10.1029/JS082i028p04351} \end{APACrefDOI}
\PrintBackRefs{\CurrentBib}

\bibitem [\protect \citeauthoryear {%
{Jain}%
\ \protect \BOthers {.}}{%
{Jain}%
\ \protect \BOthers {.}}{%
{\protect \APACyear {2015}}%
}]{%
jain2015}
\APACinsertmetastar {%
jain2015}%
\begin{APACrefauthors}%
{Jain}, S\BPBI K.%
, {Stewart}, A\BPBI I\BPBI F.%
, {Schneider}, N\BPBI M.%
, {Deighan}, J.%
, {Stiepen}, A.%
, {Evans}, J\BPBI S.%
\BDBL {}{Jakosky}, B\BPBI M.%
\end{APACrefauthors}%
\unskip\
\newblock
\APACrefYearMonthDay{2015}{}{}.
\newblock
{\BBOQ}\APACrefatitle {{The structure and variability of {M}ars upper
  atmosphere as seen in {MAVEN/IUVS} dayglow observations}} {{The structure and
  variability of {M}ars upper atmosphere as seen in {MAVEN/IUVS} dayglow
  observations}}.{\BBCQ}
\newblock
\APACjournalVolNumPages{Geophys. Res. Lett.}{42}{}{9023-9030}.
\newblock
\begin{APACrefDOI} \doi{10.1002/2015GL065419} \end{APACrefDOI}
\PrintBackRefs{\CurrentBib}

\bibitem [\protect \citeauthoryear {%
{Jakosky}%
\ \protect \BOthers {.}}{%
{Jakosky}%
\ \protect \BOthers {.}}{%
{\protect \APACyear {2015}}%
}]{%
jakosky2015}
\APACinsertmetastar {%
jakosky2015}%
\begin{APACrefauthors}%
{Jakosky}, B\BPBI M.%
, {Lin}, R\BPBI P.%
, {Grebowsky}, J\BPBI M.%
, {Luhmann}, J\BPBI G.%
, {Mitchell}, D\BPBI F.%
, {Beutelschies}, G.%
\BDBL {}{Zurek}, R.%
\end{APACrefauthors}%
\unskip\
\newblock
\APACrefYearMonthDay{2015}{}{}.
\newblock
{\BBOQ}\APACrefatitle {{The Mars Atmosphere and Volatile Evolution (MAVEN)
  Mission}} {{The Mars Atmosphere and Volatile Evolution (MAVEN)
  Mission}}.{\BBCQ}
\newblock
\APACjournalVolNumPages{Space Sci. Rev.}{}{}{}.
\newblock
\begin{APACrefDOI} \doi{10.1007/s11214-015-0139-x} \end{APACrefDOI}
\PrintBackRefs{\CurrentBib}

\bibitem [\protect \citeauthoryear {%
{Lee}%
\ \protect \BOthers {.}}{%
{Lee}%
\ \protect \BOthers {.}}{%
{\protect \APACyear {2017}}%
}]{%
lee2017}
\APACinsertmetastar {%
lee2017}%
\begin{APACrefauthors}%
{Lee}, C\BPBI O.%
, {Hara}, T.%
, {Halekas}, J\BPBI S.%
, {Thiemann}, E.%
, {Chamberlin}, P.%
, {Eparvier}, F.%
\BDBL {}{Jakosky}, B\BPBI M.%
\end{APACrefauthors}%
\unskip\
\newblock
\APACrefYearMonthDay{2017}{}{}.
\newblock
{\BBOQ}\APACrefatitle {{{MAVEN} observations of the solar cycle 24 space
  weather conditions at {M}ars}} {{{MAVEN} observations of the solar cycle 24
  space weather conditions at {M}ars}}.{\BBCQ}
\newblock
\APACjournalVolNumPages{J. Geophys. Res.}{122}{}{2768-2794}.
\newblock
\begin{APACrefDOI} \doi{10.1002/2016JA023495} \end{APACrefDOI}
\PrintBackRefs{\CurrentBib}

\bibitem [\protect \citeauthoryear {%
Ma%
, Nagy%
, Sokolov%
\BCBL {}\ \BBA {} Hansen%
}{%
Ma%
\ \protect \BOthers {.}}{%
{\protect \APACyear {2004}}%
}]{%
ma2004}
\APACinsertmetastar {%
ma2004}%
\begin{APACrefauthors}%
Ma, Y.%
, Nagy, A\BPBI F.%
, Sokolov, I\BPBI V.%
\BCBL {}\ \BBA {} Hansen, K\BPBI C.%
\end{APACrefauthors}%
\unskip\
\newblock
\APACrefYearMonthDay{2004}{}{}.
\newblock
{\BBOQ}\APACrefatitle {Three-dimensional, multispecies, high spatial resolution
  {MHD} studies of the solar wind interaction with {M}ars} {Three-dimensional,
  multispecies, high spatial resolution {MHD} studies of the solar wind
  interaction with {M}ars}.{\BBCQ}
\newblock
\APACjournalVolNumPages{J. Geophys. Res.}{109}{}{A07211, 10.1029/2003JA010367}.
\newblock
\begin{APACrefDOI} \doi{10.1029/2003JA010367} \end{APACrefDOI}
\PrintBackRefs{\CurrentBib}

\bibitem [\protect \citeauthoryear {%
{Mahaffy}%
, {Benna}%
, {Elrod}%
\BCBL {}\ \protect \BOthers {.}}{%
{Mahaffy}%
, {Benna}%
, {Elrod}%
\BCBL {}\ \protect \BOthers {.}}{%
{\protect \APACyear {2015b}}%
}]{%
mahaffy2015a}
\APACinsertmetastar {%
mahaffy2015a}%
\begin{APACrefauthors}%
{Mahaffy}, P\BPBI R.%
, {Benna}, M.%
, {Elrod}, M.%
, {Yelle}, R\BPBI V.%
, {Bougher}, S\BPBI W.%
, {Stone}, S\BPBI W.%
\BCBL {}\ \BBA {} {Jakosky}, B\BPBI M.%
\end{APACrefauthors}%
\unskip\
\newblock
\APACrefYearMonthDay{2015b}{}{}.
\newblock
{\BBOQ}\APACrefatitle {{Structure and composition of the neutral upper
  atmosphere of Mars from the {MAVEN} {NGIMS} investigation}} {{Structure and
  composition of the neutral upper atmosphere of Mars from the {MAVEN} {NGIMS}
  investigation}}.{\BBCQ}
\newblock
\APACjournalVolNumPages{Geophys. Res. Lett.}{42}{}{8951-8957}.
\newblock
\begin{APACrefDOI} \doi{10.1002/2015GL065329} \end{APACrefDOI}
\PrintBackRefs{\CurrentBib}

\bibitem [\protect \citeauthoryear {%
{Mahaffy}%
, {Benna}%
, {King}%
\BCBL {}\ \protect \BOthers {.}}{%
{Mahaffy}%
, {Benna}%
, {King}%
\BCBL {}\ \protect \BOthers {.}}{%
{\protect \APACyear {2015a}}%
}]{%
mahaffy2015}
\APACinsertmetastar {%
mahaffy2015}%
\begin{APACrefauthors}%
{Mahaffy}, P\BPBI R.%
, {Benna}, M.%
, {King}, T.%
, {Harpold}, D\BPBI N.%
, {Arvey}, R.%
, {Barciniak}, M.%
\BDBL {}{Nolan}, J\BPBI T.%
\end{APACrefauthors}%
\unskip\
\newblock
\APACrefYearMonthDay{2015a}{}{}.
\newblock
{\BBOQ}\APACrefatitle {{The Neutral Gas and Ion Mass Spectrometer on the Mars
  Atmosphere and Volatile Evolution Mission}} {{The Neutral Gas and Ion Mass
  Spectrometer on the Mars Atmosphere and Volatile Evolution Mission}}.{\BBCQ}
\newblock
\APACjournalVolNumPages{Space. Sci. Rev.}{195}{}{49-73}.
\newblock
\begin{APACrefDOI} \doi{10.1007/s11214-014-0091-1} \end{APACrefDOI}
\PrintBackRefs{\CurrentBib}

\bibitem [\protect \citeauthoryear {%
{Mendillo}%
\ \protect \BOthers {.}}{%
{Mendillo}%
\ \protect \BOthers {.}}{%
{\protect \APACyear {2011}}%
}]{%
mendillo2011}
\APACinsertmetastar {%
mendillo2011}%
\begin{APACrefauthors}%
{Mendillo}, M.%
, {Lollo}, A.%
, {Withers}, P.%
, {Matta}, M.%
, {P{\"a}tzold}, M.%
\BCBL {}\ \BBA {} {Tellmann}, S.%
\end{APACrefauthors}%
\unskip\
\newblock
\APACrefYearMonthDay{2011}{{\APACmonth{11}}}{}.
\newblock
{\BBOQ}\APACrefatitle {Modeling {M}ars' ionosphere with constraints from
  same-day observations by {M}ars {G}lobal {S}urveyor and {M}ars {E}xpress}
  {Modeling {M}ars' ionosphere with constraints from same-day observations by
  {M}ars {G}lobal {S}urveyor and {M}ars {E}xpress}.{\BBCQ}
\newblock
\APACjournalVolNumPages{J. Geophys. Res.}{116}{}{A11303,10.1029/2011JA016865}.
\newblock
\begin{APACrefDOI} \doi{10.1029/2011JA016865} \end{APACrefDOI}
\PrintBackRefs{\CurrentBib}

\bibitem [\protect \citeauthoryear {%
{Mendillo}%
, {Marusiak}%
, {Withers}%
, {Morgan}%
\BCBL {}\ \BBA {} {Gurnett}%
}{%
{Mendillo}%
\ \protect \BOthers {.}}{%
{\protect \APACyear {2013}}%
}]{%
mendillo2013}
\APACinsertmetastar {%
mendillo2013}%
\begin{APACrefauthors}%
{Mendillo}, M.%
, {Marusiak}, A\BPBI G.%
, {Withers}, P.%
, {Morgan}, D.%
\BCBL {}\ \BBA {} {Gurnett}, D.%
\end{APACrefauthors}%
\unskip\
\newblock
\APACrefYearMonthDay{2013}{}{}.
\newblock
{\BBOQ}\APACrefatitle {{A new semiempirical model of the peak electron density
  of the {M}artian ionosphere}} {{A new semiempirical model of the peak
  electron density of the {M}artian ionosphere}}.{\BBCQ}
\newblock
\APACjournalVolNumPages{Geophys. Res. Lett.}{40}{}{5361-5365}.
\newblock
\begin{APACrefDOI} \doi{10.1002/2013GL057631} \end{APACrefDOI}
\PrintBackRefs{\CurrentBib}

\bibitem [\protect \citeauthoryear {%
{Morgan}%
\ \protect \BOthers {.}}{%
{Morgan}%
\ \protect \BOthers {.}}{%
{\protect \APACyear {2008}}%
}]{%
morgan2008}
\APACinsertmetastar {%
morgan2008}%
\begin{APACrefauthors}%
{Morgan}, D\BPBI D.%
, {Gurnett}, D\BPBI A.%
, {Kirchner}, D\BPBI L.%
, {Fox}, J\BPBI L.%
, {Nielsen}, E.%
\BCBL {}\ \BBA {} {Plaut}, J\BPBI J.%
\end{APACrefauthors}%
\unskip\
\newblock
\APACrefYearMonthDay{2008}{}{}.
\newblock
{\BBOQ}\APACrefatitle {Variation of the martian ionospheric electron density
  from {M}ars {E}xpress radar soundings} {Variation of the martian ionospheric
  electron density from {M}ars {E}xpress radar soundings}.{\BBCQ}
\newblock
\APACjournalVolNumPages{J. Geophys. Res.}{113}{}{A09303, 10.1029/2008JA013313}.
\newblock
\begin{APACrefDOI} \doi{10.1029/2008JA013313} \end{APACrefDOI}
\PrintBackRefs{\CurrentBib}

\bibitem [\protect \citeauthoryear {%
{Orosei}%
\ \protect \BOthers {.}}{%
{Orosei}%
\ \protect \BOthers {.}}{%
{\protect \APACyear {2015}}%
}]{%
orosei2015}
\APACinsertmetastar {%
orosei2015}%
\begin{APACrefauthors}%
{Orosei}, R.%
, {Jordan}, R\BPBI L.%
, {Morgan}, D\BPBI D.%
, {Cartacci}, M.%
, {Cicchetti}, A.%
, {Duru}, F.%
\BDBL {}{Picardi}, G.%
\end{APACrefauthors}%
\unskip\
\newblock
\APACrefYearMonthDay{2015}{}{}.
\newblock
{\BBOQ}\APACrefatitle {{Mars {A}dvanced {R}adar for {S}ubsurface and
  {I}onospheric Sounding ({MARSIS}) after nine years of operation: {A}
  summary}} {{Mars {A}dvanced {R}adar for {S}ubsurface and {I}onospheric
  Sounding ({MARSIS}) after nine years of operation: {A} summary}}.{\BBCQ}
\newblock
\APACjournalVolNumPages{Planet. Space Sci.}{112}{}{98-114}.
\newblock
\begin{APACrefDOI} \doi{10.1016/j.pss.2014.07.010} \end{APACrefDOI}
\PrintBackRefs{\CurrentBib}

\bibitem [\protect \citeauthoryear {%
{Rishbeth}%
\ \BBA {} {Edwards}%
}{%
{Rishbeth}%
\ \BBA {} {Edwards}%
}{%
{\protect \APACyear {1989}}%
}]{%
rishbeth1989}
\APACinsertmetastar {%
rishbeth1989}%
\begin{APACrefauthors}%
{Rishbeth}, H.%
\BCBT {}\ \BBA {} {Edwards}, R.%
\end{APACrefauthors}%
\unskip\
\newblock
\APACrefYearMonthDay{1989}{}{}.
\newblock
{\BBOQ}\APACrefatitle {{The isobaric {F2}-layer}} {{The isobaric
  {F2}-layer}}.{\BBCQ}
\newblock
\APACjournalVolNumPages{J. Atmos. Terr.}{51}{}{321-338}.
\newblock
\begin{APACrefDOI} \doi{10.1016/0021-9169(89)90083-4} \end{APACrefDOI}
\PrintBackRefs{\CurrentBib}

\bibitem [\protect \citeauthoryear {%
{Rishbeth}%
, {Lyon}%
\BCBL {}\ \BBA {} {Peart}%
}{%
{Rishbeth}%
\ \protect \BOthers {.}}{%
{\protect \APACyear {1963}}%
}]{%
rishbeth1963}
\APACinsertmetastar {%
rishbeth1963}%
\begin{APACrefauthors}%
{Rishbeth}, H.%
, {Lyon}, A\BPBI J.%
\BCBL {}\ \BBA {} {Peart}, M.%
\end{APACrefauthors}%
\unskip\
\newblock
\APACrefYearMonthDay{1963}{}{}.
\newblock
{\BBOQ}\APACrefatitle {{Diffusion in the {E}quatorial {F} {L}ayer}} {{Diffusion
  in the {E}quatorial {F} {L}ayer}}.{\BBCQ}
\newblock
\APACjournalVolNumPages{J. Geophysical. Res.}{68}{}{2559-2569}.
\newblock
\begin{APACrefDOI} \doi{10.1029/JZ068i009p02559} \end{APACrefDOI}
\PrintBackRefs{\CurrentBib}

\bibitem [\protect \citeauthoryear {%
Schunk%
\ \BBA {} Nagy%
}{%
Schunk%
\ \BBA {} Nagy%
}{%
{\protect \APACyear {2009}}%
}]{%
schunk2009}
\APACinsertmetastar {%
schunk2009}%
\begin{APACrefauthors}%
Schunk, R\BPBI W.%
\BCBT {}\ \BBA {} Nagy, A\BPBI F.%
\end{APACrefauthors}%
\unskip\
\newblock
\APACrefYear{2009}.
\newblock
\APACrefbtitle {Ionospheres} {Ionospheres}\ (\PrintOrdinal{Second}\ \BEd).
\newblock
\APACaddressPublisher{New York}{Cambridge University Press}.
\PrintBackRefs{\CurrentBib}

\bibitem [\protect \citeauthoryear {%
Thiemann%
, Andersson%
\BCBL {}\ \protect \BOthers {.}}{%
Thiemann%
, Andersson%
\BCBL {}\ \protect \BOthers {.}}{%
{\protect \APACyear {2018a}}%
}]{%
thiemann2017a}
\APACinsertmetastar {%
thiemann2017a}%
\begin{APACrefauthors}%
Thiemann, E\BPBI M\BPBI B.%
, Andersson, L.%
, Lillis, R.%
, Withers, P.%
, Xu, S.%
, Elrod, M.%
\BDBL {}Deighan, J.%
\end{APACrefauthors}%
\unskip\
\newblock
\APACrefYearMonthDay{2018a}{}{}.
\newblock
{\BBOQ}\APACrefatitle {{The Mars Topside Ionosphere Response to the {X}8.2
  Solar Flare of 10 {S}eptember 2017}} {{The Mars Topside Ionosphere Response
  to the {X}8.2 Solar Flare of 10 {S}eptember 2017}}.{\BBCQ}
\newblock
\APACjournalVolNumPages{Geophys. Res. Lett.}{45}{}{}.
\newblock
\begin{APACrefDOI} \doi{10.1029/2018GL077730} \end{APACrefDOI}
\PrintBackRefs{\CurrentBib}

\bibitem [\protect \citeauthoryear {%
{Thiemann}%
\ \protect \BOthers {.}}{%
{Thiemann}%
\ \protect \BOthers {.}}{%
{\protect \APACyear {2017}}%
}]{%
thiemann2017}
\APACinsertmetastar {%
thiemann2017}%
\begin{APACrefauthors}%
{Thiemann}, E\BPBI M\BPBI B.%
, {Chamberlin}, P\BPBI C.%
, {Eparvier}, F\BPBI G.%
, {Templeman}, B.%
, {Woods}, T\BPBI N.%
, {Bougher}, S\BPBI W.%
\BCBL {}\ \BBA {} {Jakosky}, B\BPBI M.%
\end{APACrefauthors}%
\unskip\
\newblock
\APACrefYearMonthDay{2017}{}{}.
\newblock
{\BBOQ}\APACrefatitle {{The {MAVEN} {EUVM} model of solar spectral irradiance
  variability at {M}ars: {A}lgorithms and results}} {{The {MAVEN} {EUVM} model
  of solar spectral irradiance variability at {M}ars: {A}lgorithms and
  results}}.{\BBCQ}
\newblock
\APACjournalVolNumPages{J. Geophys. Res.}{122}{}{2748-2767}.
\newblock
\begin{APACrefDOI} \doi{10.1002/2016JA023512} \end{APACrefDOI}
\PrintBackRefs{\CurrentBib}

\bibitem [\protect \citeauthoryear {%
Thiemann%
, Eparvier%
\BCBL {}\ \protect \BOthers {.}}{%
Thiemann%
, Eparvier%
\BCBL {}\ \protect \BOthers {.}}{%
{\protect \APACyear {2018b}}%
}]{%
thiemann2018}
\APACinsertmetastar {%
thiemann2018}%
\begin{APACrefauthors}%
Thiemann, E\BPBI M\BPBI B.%
, Eparvier, F\BPBI G.%
, Bougher, S\BPBI W.%
, Dominique, M.%
, Andersson, L.%
, Girazian, Z.%
\BDBL {}Jakosky, B\BPBI M.%
\end{APACrefauthors}%
\unskip\
\newblock
\APACrefYearMonthDay{2018b}{}{}.
\newblock
{\BBOQ}\APACrefatitle {Mars {T}hermospheric {V}ariability {R}evealed by {MAVEN}
  {EUVM} {S}olar {O}ccultations: {S}tructure at {A}phelion and {P}erihelion and
  {R}esponse to {EUV} {F}orcing} {Mars {T}hermospheric {V}ariability {R}evealed
  by {MAVEN} {EUVM} {S}olar {O}ccultations: {S}tructure at {A}phelion and
  {P}erihelion and {R}esponse to {EUV} {F}orcing}.{\BBCQ}
\newblock
\APACjournalVolNumPages{J. Geophys. Res.}{123}{}{}.
\newblock
\begin{APACrefDOI} \doi{10.1029/2018JE005550} \end{APACrefDOI}
\PrintBackRefs{\CurrentBib}

\bibitem [\protect \citeauthoryear {%
{Vogt}%
\ \protect \BOthers {.}}{%
{Vogt}%
\ \protect \BOthers {.}}{%
{\protect \APACyear {2017}}%
}]{%
vogt2017}
\APACinsertmetastar {%
vogt2017}%
\begin{APACrefauthors}%
{Vogt}, M\BPBI F.%
, {Withers}, P.%
, {Fallows}, K.%
, {Andersson}, L.%
, {Girazian}, Z.%
, {Mahaffy}, P\BPBI R.%
\BDBL {}{Jakosky}, B\BPBI M.%
\end{APACrefauthors}%
\unskip\
\newblock
\APACrefYearMonthDay{2017}{}{}.
\newblock
{\BBOQ}\APACrefatitle {{{MAVEN} observations of dayside peak electron densities
  in the ionosphere of {M}ars}} {{{MAVEN} observations of dayside peak electron
  densities in the ionosphere of {M}ars}}.{\BBCQ}
\newblock
\APACjournalVolNumPages{J. Geophys. Res.}{122}{}{891-906}.
\newblock
\begin{APACrefDOI} \doi{10.1002/2016JA023473} \end{APACrefDOI}
\PrintBackRefs{\CurrentBib}

\bibitem [\protect \citeauthoryear {%
{Withers}%
}{%
{Withers}%
}{%
{\protect \APACyear {2009}}%
}]{%
withers2009a}
\APACinsertmetastar {%
withers2009a}%
\begin{APACrefauthors}%
{Withers}, P.%
\end{APACrefauthors}%
\unskip\
\newblock
\APACrefYearMonthDay{2009}{}{}.
\newblock
{\BBOQ}\APACrefatitle {A review of observed variability in the dayside
  ionosphere of {M}ars} {A review of observed variability in the dayside
  ionosphere of {M}ars}.{\BBCQ}
\newblock
\APACjournalVolNumPages{Adv. Space Res.}{44}{}{277-307}.
\newblock
\begin{APACrefDOI} \doi{10.1016/j.asr.2009.04.027} \end{APACrefDOI}
\PrintBackRefs{\CurrentBib}

\bibitem [\protect \citeauthoryear {%
{Withers}%
, {Vogt}%
, {Mahaffy}%
\BCBL {}\ \protect \BOthers {.}}{%
{Withers}%
, {Vogt}%
, {Mahaffy}%
\BCBL {}\ \protect \BOthers {.}}{%
{\protect \APACyear {2015}}%
}]{%
withers2015a}
\APACinsertmetastar {%
withers2015a}%
\begin{APACrefauthors}%
{Withers}, P.%
, {Vogt}, M.%
, {Mahaffy}, P.%
, {Benna}, M.%
, {Elrod}, M.%
\BCBL {}\ \BBA {} {Jakosky}, B.%
\end{APACrefauthors}%
\unskip\
\newblock
\APACrefYearMonthDay{2015}{}{}.
\newblock
{\BBOQ}\APACrefatitle {{Changes in the thermosphere and ionosphere of {M}ars
  from V{}iking to {MAVEN}}} {{Changes in the thermosphere and ionosphere of
  {M}ars from V{}iking to {MAVEN}}}.{\BBCQ}
\newblock
\APACjournalVolNumPages{Geophys. Res. Lett.}{42}{}{9071-9079}.
\newblock
\begin{APACrefDOI} \doi{10.1002/2015GL065985} \end{APACrefDOI}
\PrintBackRefs{\CurrentBib}

\bibitem [\protect \citeauthoryear {%
{Withers}%
, {Vogt}%
, {Mayyasi}%
\BCBL {}\ \protect \BOthers {.}}{%
{Withers}%
, {Vogt}%
, {Mayyasi}%
\BCBL {}\ \protect \BOthers {.}}{%
{\protect \APACyear {2015}}%
}]{%
withers2015}
\APACinsertmetastar {%
withers2015}%
\begin{APACrefauthors}%
{Withers}, P.%
, {Vogt}, M.%
, {Mayyasi}, M.%
, {Mahaffy}, P.%
, {Benna}, M.%
, {Elrod}, M.%
\BDBL {}{Jakosky}, B.%
\end{APACrefauthors}%
\unskip\
\newblock
\APACrefYearMonthDay{2015}{}{}.
\newblock
{\BBOQ}\APACrefatitle {{Comparison of model predictions for the composition of
  the ionosphere of {M}ars to {MAVEN} {NGIMS} data}} {{Comparison of model
  predictions for the composition of the ionosphere of {M}ars to {MAVEN}
  {NGIMS} data}}.{\BBCQ}
\newblock
\APACjournalVolNumPages{Geophys. Res. Letter}{42}{}{8966-8976}.
\newblock
\begin{APACrefDOI} \doi{10.1002/2015GL065205} \end{APACrefDOI}
\PrintBackRefs{\CurrentBib}

\bibitem [\protect \citeauthoryear {%
{Wood}%
\ \BBA {} {Paige}%
}{%
{Wood}%
\ \BBA {} {Paige}%
}{%
{\protect \APACyear {1992}}%
}]{%
wood1992}
\APACinsertmetastar {%
wood1992}%
\begin{APACrefauthors}%
{Wood}, S\BPBI E.%
\BCBT {}\ \BBA {} {Paige}, D\BPBI A.%
\end{APACrefauthors}%
\unskip\
\newblock
\APACrefYearMonthDay{1992}{}{}.
\newblock
{\BBOQ}\APACrefatitle {{Modeling the {M}artian seasonal {CO2} cycle. {I} -
  {F}itting the {V}iking {L}ander pressure curves. {II} - {I}nterannual
  variability}} {{Modeling the {M}artian seasonal {CO2} cycle. {I} - {F}itting
  the {V}iking {L}ander pressure curves. {II} - {I}nterannual
  variability}}.{\BBCQ}
\newblock
\APACjournalVolNumPages{Icarus}{99}{}{1-27}.
\newblock
\begin{APACrefDOI} \doi{10.1016/0019-1035(92)90166-5} \end{APACrefDOI}
\PrintBackRefs{\CurrentBib}

\end{thebibliography}




%

\newpage

\newpage

\newpage

\newpage

\newpage

 \newpage

	\newpage

\newpage

\end{document}